\documentclass{autart}

\usepackage{graphics} % for pdf, bitmapped graphics files
\usepackage{subcaption}
\usepackage{verbatim}
\usepackage{epsfig}
\usepackage{amsmath}
\usepackage[justification=centering]{caption}
\usepackage{amssymb}
\usepackage{psfrag}
\usepackage[T1]{fontenc}
\usepackage{dsfont}
\usepackage{mathtools}
\DeclarePairedDelimiter{\ceil}{\lceil}{\rceil}
\usepackage{booktabs}
\usepackage[export]{adjustbox}

\usepackage[norelsize,ruled]{algorithm2e}
\usepackage{float}
\usepackage{textcomp}
\usepackage{xcolor}
\usepackage{etoolbox}

\usepackage[noadjust]{cite}

\newtheorem{theorem}{Theorem}[section]
\newtheorem{lemma}{Lemma}[section]

\newtheorem{definition}{Definition}
\newtheorem{assumption}{Assumption}
\newtheorem{proposition}{Proposition}[section]

\newtheorem{remark}{Remark}

\definecolor{cadmiumgreen}{rgb}{0.0, 0.42, 0.24}

\DeclareMathOperator*{\minimize}{minimize}
\DeclareMathOperator*{\dom}{dom}

\DeclareMathOperator*{\x}{x}
\DeclareMathOperator*{\z}{z}

\begin{document}

\begin{frontmatter}
% \runtitle{Finite-time Approximate Consensus based Distributed Optimization on Directed graphs}  % Running title for regular 

\title{Gradient-Consensus: Linearly Convergent Distributed Optimization Algorithm over Directed Graphs\thanksref{footnoteinfo}} % Title, preferably not more 

\thanks[footnoteinfo]{\tiny{This work was authored in part by NREL, operated by Alliance for Sustainable Energy, LLC, for the U.S. Department of Energy (DOE) under Contract No. DE-AC36-08GO28308. Funding provided by the Advanced Research Projects Agency-Energy under Grant DE-AR0001016. The views expressed in the article do not necessarily represent the views of the DOE or the U.S. Government. The U.S. Government retains and the publisher, by accepting the article for publication, acknowledges that the U.S. Government retains a nonexclusive, paid-up, irrevocable, worldwide license to publish or reproduce the published form of this work, or allow others to do so, for U.S. Government purposes.}}
\noindent \author[Minneapolis]{Vivek Khatana}\ead{khata010@umn.edu},               % e-mail address 
\author[NREL]{Govind Saraswat}\ead{govind.saraswat@nrel.gov}, 
\author[Minneapolis]{Sourav Patel}\ead{patel292@umn.edu},  % (ead) as shown
\author[Minneapolis]{Murti V. Salapaka}\ead{murtis@umn.edu}
\address[Minneapolis]{Department of Electrical and Computer Engineering, University of Minnesota, Minneapolis, USA}  % Please supply
\address[NREL]{National Renewable Energy Laboratory (NREL), Golden, CO, USA}    


\begin{abstract}
In this article, we propose a new approach, {\it optimize then agree} for minimizing a sum, $f = \sum_{i=1}^n f_i$, of convex objective functions $f_i$'s, where, $f_i$ is only available locally to the agent $i$, over a directed graph of $n$ agents. The \textit{optimize then agree} approach decouples the optimization step and the consensus step in a distributed optimization framework. One of the key motivations for \textit{optimize then agree} is to guarantee that the disagreement between the estimates of the agents during every iteration of the distributed optimization algorithm remains under any \textit{apriori} specified tolerance; existing algorithms do not provide such a guarantee which is required in many practical scenarios. In this method, each agent during each iteration maintains an estimate of the optimal solution and, utilizes its locally available gradient information along with a finite-time approximate consensus protocol to move towards the optimal solution (hence the name Gradient-Consensus algorithm). We establish that the proposed algorithm has a global R-linear rate of convergence if the aggregate function $f$ is strongly convex and Lipschitz differentiable. We also show that under the relaxed assumption of $f_i$'s being convex and Lipschitz differentiable, the objective function error residual decreases at a Q-linear rate (in terms of the number of gradient computation steps) until it reaches a small value, which can be managed using the tolerance value specified on the finite-time approximate consensus protocol; no existing method in the literature has such strong convergence guarantees when $f_i$ are not necessarily strongly convex functions. The communication overhead for the improved  guarantees on meeting constraints and better convergence of our algorithm is $O(k\log k)$ iterates in comparison to $O(k)$ of the traditional algorithms. Further, we numerically evaluate the performance of the proposed algorithm by solving a distributed logistic regression problem. 
\end{abstract}

\end{frontmatter}

\begin{section}{Introduction}\label{sec:introduction}
In this article, we consider a group of $n$ agents connected as a directed graph, with a goal of solving the following distributed optimization problem:
\begin{align}\label{eq:introprob}
   \textstyle \minimize\limits_{x \in \mathbb{R}^p} \ f(x) =  \sum_{i=1}^{n} f_i(x),
\end{align}
where, $x \in \mathbb{R}^p$ is a global optimization variable, and each function $f_i$ : $\mathbb{R}^p \rightarrow \mathbb{R}$ is a convex cost function known only to agent $i$. Due to the underlying directed interconnection structure,  agents can only send (receive) information to (from) their neighboring nodes connected via a unidirectional link in the directed communication structure. 
% Various problems arising in multi-agent coordination and control \cite{olshevsky2010efficient}, robotics \cite{ren2007information}, sensor networks \cite{rabbat2004distributed}, wireless systems \cite{predd2006distributed}, machine learning \cite{predd2009collaborative} can be posed as problem (\ref{eq:introprob}).
The study of the distributed optimization problem initiated with the seminal works \cite{tsitsiklis1984problems}, \cite{bertsekas1989parallel}. Since then, numerous algorithms to minimize the sum of convex functions in a distributed manner are devised; see \cite{nedic2009distributed, lobel2010distributed, nedic2010asynchronous, duchi2011dual, chen2012diffusion}. Most of the existing first-order methods for solving~(\ref{eq:introprob}) include the distributed gradient descent method \cite{nedic2009distributed}, \cite{yuan2016convergence} and its modifications \cite{nedic2014distributed}, \cite{nedic2016stochastic}. The authors in \cite{shi2015extra} proposed an exact first order method that converges to optimal solution while using a fixed step-size in the gradient updates. The works in \cite{shi2015proximal}, \cite{li2019decentralized} have developed algorithms based on proximal-gradient to tackle~(\ref{eq:introprob}) with proximal friendly $f_i$'s. Most existing works including, \cite{nedic2009distributed, duchi2011dual, yuan2016convergence,  nedic2014distributed, shi2015extra,olshevsky2017linear, shi2015proximal, li2019decentralized }, impose restrictive conditions such as requirement of doubly-stochastic weight matrices and need of balanced undirected graphs. The first work to propose a distributed optimization method for directed graphs appeared in \cite{tsianos2012push}. Subsequently, the authors in \cite{nedic2014distributed} incorporated the push-sum protocol \cite{kempe2003gossip} with an earlier Distributed Gradient Descent (DGD) method \cite{nedic2009distributed} to propose the subgradient-push algorithm for time-varying directed graphs.
% To get improved convergence rates over directed networks, \cite{xi2017dextra}, \cite{zeng2015extrapush} proposed linearly convergent schemes for strongly convex and smooth  $f_i$'s by modifying \cite{shi2015extra} using the push-sum technique.
Recent works \cite{xin2018linear} and \cite{pu2020push} proposed distributed optimization schemes for directed networks that make use of both row and column stochastic matrices in one iteration of the algorithm. 
\textcolor{black}{ While there is a proliferating literature on developing distributed
optimization methods for problem~(\ref{eq:introprob}), most of them suffer from mismatch between the estimates of the agents at any finite time termination of the algorithm. As an illustration, we consider, an equivalent reformulation of problem~(\ref{eq:introprob}):
\begin{align}\label{eq:prob_cons_constraints}
  \textstyle \minimize\limits_{x^1,\dots,x^n} & \textstyle \ \sum_{i=1}^{n} f_i(x^i)\\
  \mbox{subject to} \ x^i = & \ x^j, \  \mbox{for all} \ i,j \in \mathbf{V}, \nonumber
\end{align}
where, $x^i, x^j \in \mathbb{R}^p$ are local estimates (of the optimal solution) of agent $i$ and $j$ respectively and the equality constraints enforce agreement between the local estimates of the agents. The design of existing algorithms that focus on finding the solution of~(\ref{eq:prob_cons_constraints}) emphasize solving the unconstrained problem and reach on agreement between the agents estimates only asymptotically. \textcolor{black}{In particular, at the intermediate iterations of the algorithm, the agents' estimates often allow considerable mismatch and hence, do not provide a practically feasible solution to problem~(\ref{eq:prob_cons_constraints}). The disagreement of agents' estimates can be problematic for practical applications where obtaining a feasible solution with guarantees in finite-time is crucial for adequate performance and stability of the system.} In order to motivate the discussion, we next provide illustrative examples of real-world cyber-physical systems where agreement on a solution is imperative and failing to do so can result in undesirable consequences to the system.
\begin{itemize}
\item[i)] \textcolor{black}{\textit{Economic Dispatch in power systems}:  The economic dispatch (ED) is an optimization problem that tries to minimize the cost of dispatching $n$ generation sources to meet the total load demand $P_\ell$ while meeting generation constraints in the power system. Let the cost of operation for a generation source $i$ be given as, $C_i(P_i) = \alpha_i + \beta_i P_i + \gamma_i P_i^2$, where $P_i$ is the amount of power generated by source $i$ \cite{wood2013power}. The ED problem can be formulated as follows: 
    \begin{align*}
    \minimize\limits_{P_1,\dots,P_n}  & \ \textstyle \sum_{i=1}^n C_i(P_i)\\
    \mbox{subject to} & \ \textstyle \sum_{i=1}^n P_{i} = P_{\ell},\\
       &P_{i}^{min} \leq P_{i} \leq P_{i}^{max} \ \  \mbox{for} \ i = 1,2,\dots,n,
\end{align*} 
where, $P_{i}^{min}$ and $P_{i}^{max}$ are the minimum and maximum power rating of generator $i$.  The ED problem
can be solved in a centralized manner by the Lagrange multiplier method \cite{wood2013power}. By introducing the Lagrange multiplier $\lambda$, the solution of the ED problem
can be obtained by solving the following equations:
\begin{align}\label{eq:EDlambda}
   \textstyle \frac{\text{d}C_i(P_i)}{\text{d}P_i} = \lambda, \forall i.
\end{align}
It is important while solving the ED problem distributively to ensure that the Lagrangian multiplier for all the generation sources should agree as required by~(\ref{eq:EDlambda}). This is also referred to as the \textit{equal incremental cost optimization criterion} in power systems \cite{wood2013power} which ensures that the cost of increasing generation anywhere in the power network is same for all the distributed computational agents solving the economic dispatch problem. Let $q_i = \frac{\text{d}C_i(P_i)}{\text{d}P_i} = 2\gamma_i P_i + \beta_i$. Using the definition of $q_i$ and~(\ref{eq:EDlambda}), solution to the ED problem can be obtained by solving the following equivalent distributed optimization problem: 
\begin{align*}
    \minimize_{q_1, \dots, q_n} & \ \ \textstyle  \big|\sum_{i=1}^n \frac{q_i - \beta_i}{2 \gamma_i} -  P_{\ell}\big|\\
    \mbox{subject to} & \ \ q_i = q_j, \forall i,j\\
       & \ \ q_{i}^{min} \leq q_{i} \leq q_{i}^{max} \ \  \forall i,
\end{align*}
where, $q^{min}_i = 2\gamma_i P^{min}_i + \beta_i$ and $q^{max}_i = 2\gamma_i P^{max}_i + \beta_i$. The ED problem needs to be solved fast as the solutions obtained are used to dispatch generation sources in a real-time electricity market as power system states change rapidly \cite{patel2020distributed}. This imposes a restriction on the available computation time and the distributed algorithm solving the ED problem has to provide estimates of the solution (even sub-optimal) in a short amount of time. Hence, it is imperative that the distributed optimization algorithm maintain constraint feasibility during the iterations, as an infeasible dispatch solution can result in economic and stability issues (such as large frequency deviations) in the power system network.}
\item [ii)] \textcolor{black}{\textit{Rendezvous of Multi-Robot Systems:} Consider, a group of $n$ mobile robots. The control objective for the robots is to meet at a common agreed upon location $p^*$. We use $p_i$ and $t_i$ to denote the estimate of the position $p^*$ and the time at which robot $i$ considers the group should meet. It is desired to develop a distributed algorithm such that these estimates of meeting time and position should be same for all the robots, i.e., $p_i = p_j$ and $t_i = t_j$ for all $i,j$ \cite{Rendezvous_example}. The robot rendezvous problem can be formulated as the following distributed optimization problem:
\begin{align*}
    \minimize_{(p_1, t_1), \dots, (p_n,t_n)} & \ \textstyle \sum_{i=1}^n f_i(p_i,t_i) \\
    \mbox{subject to} & \ \ p_i = p_j , t_i = t_j, \  \forall i \\
    & \ \ u_i(p_i,t_i) \leq 0  \ \forall i,
\end{align*}
where, $f_i$ is an objective function of robot $i$ (for example, a function to calculate the trajectory of robot $i$) and $u_i(p_i,t_i)$ is the constraint function of robot $i$ (for example, constraints on local battery power usage). Note that preserving the spatiotemporal connectivity among the agents is an important requirement in robot systems \cite{Rendezvous_example},\cite{fax2002information}. Hence, while solving the rendezvous problem it is critical to maintain an agreement about the position of each robot during the intermediate iterations of the algorithm.}
\end{itemize}
In this article, we introduce a new framework for designing distributed optimization algorithms to solve~(\ref{eq:prob_cons_constraints}). Our approach is motivated towards reducing the inaccuracy of the consensus step in the existing distributed optimization algorithms in the literature. In particular, we propose an \textit{optimize then agree} framework to decouple the gradient-descent step from the consensus step, used in tandem in most existing distributed-optimization algorithms (see \cite{nedic2009distributed, yuan2016convergence, shi2015extra, pu2020push} for example) to reduce the mismatch between the estimates maintained by different agents during each iteration of the algorithm. Based on the \textit{optimize then agree} framework we develop a novel first order distributed optimization method, termed as Gradient-Consensus (abbreviated as \texttt{GradConsensus}) algorithm for directed graph topologies. Each iteration of the \texttt{GradConsensus} algorithm comprises two steps: a local gradient-descent step at each agent followed by a finite-time approximate consensus protocol. The finite-time approximate consensus protocol is designed such that after the consensus step the updates of all agents are $\varepsilon$-close to each other, where $\varepsilon$ is a parameter independent of the problem data and can be chosen to get a desired level of accuracy. We remark here that a few earlier works \cite{jakovetic2014fast, chen2012fast,johansson2008subgradient}
% \cite{jakovetic2014fast, chen2012fast,johansson2008subgradient, berahas2018balancing}
have explored the idea of utilizing multiple consensus steps. The article \cite{johansson2008subgradient} does not provide any convergence rate estimates for the proposed scheme. The protocols in \cite{jakovetic2014fast, chen2012fast,johansson2008subgradient} depend on a symmetric double weight matrix, the synthesis of which requires global information of the network making them unsuitable to be implemented in directed networks. The scheme proposed in this article is amenable to directed networks and does not rely on doubly stochastic matrices; thus the synthesis of the algorithm and it's implementation do not require centralized information of the communication network. Moreover, we rigorously establish convergence rate estimates for the proposed algorithm.
}
The main contributions of this article are as follows:
\textcolor{black}{
\begin{enumerate}
    \item We focus on the ``disagreement'' between the estimates of different agents in the existing distributed optimization algorithms which we call the consensus constraint violation. We demonstrate consensus constraint violation for three existing algorithms in the literature while solving a distributed logistic regression problem (refer Section 4). \textcolor{black}{For any finite iteration, the solution obtained by the algorithms do not provide a feasible solution of the original distributed optimization problem. For applications (discussed earlier) where getting a feasible solution is critical during every time-instant of operation the existing algorithms do not provide a viable solution.} To address this issue, we present a novel \textit{optimize then agree} framework with a motivation of decoupling the consensus and optimization steps used in tandem in existing distributed optimization schemes. Our framework provides a new perspective on solving distributed optimization problems and presents guidelines for the development of new algorithms. Based on this framework we develop a first-order distributed algorithm termed as \texttt{GradConsensus} to solve distributed optimization problems over directed graph topologies.
    \item The \texttt{GradConsensus} algorithm is suitable for directed graphs unlike most of the existing works in the literature. It utilizes \textbf{only} the knowledge of the out-degree of each agent in the graph and is based on a column stochastic matrix which is amenable for distributed synthesis scenarios where full information of the network connectivity structure is not available and creating a doubly-stochastic matrix is not tractable.
    \item We analyze the convergence of the \texttt{GradConsensus} algorithm under two scenarios:
    \begin{enumerate}
        \item We establish a global R-linear rate of convergence for \texttt{GradConsensus} in terms of the number of gradient computations performed by the algorithm under the assumption of aggregate function $f$ being strongly-convex and Lipschitz differentiable. 
        \item We show that under the relaxed assumption of individual functions $f_i$ being convex and Lipschitz differentiable, the objective function residual (defined later in the article) under the \texttt{GradConsensus} algorithm converges at a Q-linear rate (in terms of the number of gradient computations) until reaching a small $O(\varepsilon_0)$ value, where $\varepsilon_0$ depends on the tolerance of the finite-time consensus protocol. We remark that this stronger convergence guarantee, compared to the existing literature, under the assumption that $f_i$ are not necessarily strongly convex is a novel result. The improved guarantees accrue with an overhead of $O(k\log k)$ (with objectives having uniformly  bounded derivatives) in contrast to $O(k)$ communication steps by the $k^{th}$ iterate for existing approaches.
    \end{enumerate}
    \item \textcolor{black}{We present numerical tests comparing the performance of the proposed \texttt{GradConsensus} in solving the distributed logistic regression problem to existing algorithms in the literature. The numerical simulations demonstrate that the proposed \texttt{GradConsensus} algorithm requires significantly less number of gradient calculations compared to the other algorithms and in applications where the cost of gradient computation is large our algorithm provides a better solution. We demonstrate the consensus constraint violation during the intermediate iterations in other algorithms in the literature.}
\end{enumerate}
}

A preliminary work on Gradient-Consensus by the authors can be found in \cite{khatana2020acc}. In this article, we significantly extend our earlier work by introducing the \textit{optimize then agree} framework. We present theorectical convergence guarantees for the \texttt{GradConsensus} algorithm, account quantitatively for number of communication steps required, provide new theoretical results and provide illustrative examples to corroborate the theoretical analysis. Such work is not present in the preliminary conference work. 

The rest of the paper is organized as follows: Subsection~\ref{sec:defn} provides the definitions and notations used in the article. In Section~\ref{sec:gradcons}, we present the \textit{optimize then agree} framework and the proposed \texttt{GradConsensus} algorithm along with discussion on its design and supporting results. We establish the convergence guarantees for the \texttt{GradConsensus} algorithm under two different set of assumptions in Section~\ref{sec:convgAnalysis}. In Section~\ref{sec:simulations}, we demonstrate the resulting disagreement between the agents' estimates of the solution generated by three existing algorithms in the literature while solving a distributed logistic regression problem. Further, we compare the performance of the \texttt{GradConsensus} algorithm in solving the distributed logistic regression problem with the other existing algorithms in the literature. Section~\ref{sec:conclusion} provides the conclusion.

\begin{subsection}{Definitions and Notations}\label{sec:defn}

\begin{definition}(Directed Graph)
A directed graph $\mathbf{G}$ is a pair $(\mathbf{V},\mathbf{E})$ where $\mathbf{V}$ is a set of vertices (or nodes) and $\mathbf{E}$ is a set of edges, which are ordered subsets of two distinct elements of $\mathbf{V}$. If an edge from $j \in \mathbf{V}$ to $i \in \mathbf{V}$ exists then it is denoted as $(i,j)\in \mathbf{E}$. 
\end{definition} 

\begin{definition}(Path) 
In a directed graph, a directed path from node $m$ to $\ell$
exists if there is a sequence of distinct directed edges of $\mathbf{G}$ of the form $(d_{1},m),(d_{2},d_{1}),...,(\ell,d_{\ell}).$
\end{definition}

\begin{definition}(Strongly Connected Graph) A directed graph is strongly connected if and only if it has a directed path between each pair of distinct nodes $k$ and $\ell$. 
\end{definition}

\begin{definition}(Diameter of a Graph)
The diameter of a directed graph $\mathbf{G}$ is the longest shortest directed path between any two nodes of $\mathbf{G}$. 
\end{definition}

\begin{definition}(In-Neighborhood) The set $N^{in}_i = \{j \ | \ (i,j)\in \mathbf{E}\}$ of in-neighbors of node $i \in \mathbf{V}$ not including the node $i$, is called the in-neighborhood of node $i$ with $|N^{in}_i|$ denoting the number of in-neighbors (in-degree) of node $i$.
\end{definition}

\begin{definition}(Out-Neighborhood) The set $N^{out}_i = \{j \ | \ (j,i)\in \mathbf{E}\}$ of out-neighbors of node $i \in \mathbf{V}$ not including the node $i$, is called the out-neighborhood of node $i$ with $|N^{out}_i|$ denoting the number of out-neighbors (out-degree) of node $i$. 
\end{definition}

\begin{definition}(Column Stochastic Matrix) A $n\times n$ real matrix $\mathbf{P}=[p_{ij}]$
is column-stochastic matrix if $\sum_{i=1}^{n}p_{ij}=1$ where, $0 \leq p_{ij}\leq 1$, for $1\leq i,j\leq n$. 
\end{definition}

\begin{definition}(Irreducible Matrix) A $n \times n$ matrix $\mathbf{P}$ is irreducible if for all $i,j$, there exist $m \in \mathbb{N}$ such that $[\mathbf{P}^m]_{ij} > 0$.
\end{definition}

\begin{definition}(Primitive Matrix) A non-negative matrix $\mathbf{P}$ is primitive if it is irreducible and has only one eigenvalue of maximum modulus. 
\end{definition}

\begin{definition}(Lipschitz Differentiability)
A differentiable function $f$ is called Lipschitz differentiable with constant $L$, if there exists $L>0$ such that the following inequality holds:  
\begin{align*}
    \| \nabla f(x) - \nabla f(y)\| \leq L \|x-y\|, \ \forall \ x,y \in \dom f.
\end{align*}

\end{definition}
\begin{definition}(Strongly Convex Function) A differentiable function $f$ is called strongly convex with parameter $\sigma$, if there exists $\sigma > 0$ such that the following inequality holds for all $x, y$ in the domain of $f$: 
\begin{align*}
   \langle \nabla f(x)-\nabla f(y), x-y \rangle \geq \sigma \|x-y\|^2, \ \forall \ x,y \in \dom f.
\end{align*}
\end{definition}
\end{subsection}

\noindent Each agent $i \in \mathbf{V}$ maintains a local estimate $x^i(k) := [x^i_1(k) \dots x^i_p(k)]  \in \mathbb{R}^p$ at iteration $k$ of the \texttt{GradConsensus} algorithm. Let, $\mathcal{X}^* = \{x \in \mathbb{R}^p | \sum_{i=1}^{n} f_i(x) = f^*\}$ denote the set of solutions to~(\ref{eq:introprob}), with $f^*$ being the optimal objective function value. We use a directed graph $\mathbf{G}(\mathbf{V},\mathbf{E})$ with $n$ nodes, to model the network interconnection between the $n$ agents and define $\mathbf{D}$ to be an upper bound on the diameter of the graph $\mathbf{G}(\mathbf{V},\mathbf{E})$. \textcolor{black}{Throughout the article, we will use $\|x\|$ to denote the 2-norm of the vector $x \in \mathbb{R}^p$ unless stated otherwise. Further, we use the notation $\ceil*.$ to denote the least integer function or the ceiling function, defined as: 
given $x \in \mathbb{R}, \ceil*x = \min \{ m \in \mathbb{Z} | m \geq x\},$ where $\mathbb{Z}$ is the set of integers.}
\end{section}

\begin{section}{The Optimize then Agree Framework }\label{sec:gradcons}
\textcolor{black}{
% Consider,the problem in~(\ref{eq:prob_cons_constraints}) 
% an equivalent reformulation of problem~(\ref{eq:introprob}):
% \begin{align}\label{eq:prob_cons_constraints}
%   \textstyle \minimize\limits_{x \in \mathbb{R}^p} & \textstyle \sum_{i=1}^{n} f_i(x^i)\\
%   \mbox{subject to} \ x^i = & \ x^j, \  \mbox{for all} \ i,j \in \mathbf{V}, \nonumber
% \end{align}
% where, $x^i, x^j \in \mathbb{R}^p$ are local estimates (of the optimal solution) of agent $i$ and $j$ respectively and the equality constraints enforce agreement between the local estimates of the agents.
In this section, we focus on problem~(\ref{eq:prob_cons_constraints}) and present the proposed \textit{optimize then agree} framework. We emphasize that the proposed framework can be applied to many existing algorithms in the literature that utilize an optimization scheme in tandem with a consensus step for the state updates. Here we take the example of the classical DGD method \cite{nedic2009distributed} for explanation. During the iteration $k$ of DGD every agent $i$ updates its local estimate $x^i(k) \in \mathbb{R}^p$ as follows: 
\begin{align}\label{eq:dgd}
   x^i(k) = \textstyle \sum_{j \in\mathit{N^{in}_{i}} \cup {i}}p_{ij} x^j(k-1) - \alpha \nabla f_i (x^i(k-1)), 
\end{align}
where, the weights $ 0 \leq p_{ij} \leq 1$ are such that $\sum_j p_{ij} = \sum_i p_{ij} = 1$. The first term $\sum_{j \in\mathit{N^{in}_{i}} \cup {i}}p_{ij} x^j(k-1)$ in~(\ref{eq:dgd}) corresponds to a local consensus step and the second term $-\alpha \nabla f_i(x^i(k-1))$ denotes a local gradient step. Notice that due to a local consensus step the update~(\ref{eq:dgd}) does not account for the agreement between the estimates over the entire network of agents during iteration $k$. If update~(\ref{eq:dgd}) is terminated after some finite number of iterations the solution estimates will have disagreement and will not satisfy the equality constraints in problem~(\ref{eq:prob_cons_constraints}) and hence the (finite-time) solution generated will not be feasible for the original problem. To address this issue we propose the \textit{optimize then agree} framework where the optimization step (gradient-descent in this case) and the consensus scheme are decoupled from each other. The optimize then agree framework enforces that the mismatch between the agent estimates remain below any specified threshold (and hence manages the consensus constraint violation). Here, each agent employs an optimization scheme to steer towards the solution of its local sub-problem and utilizes a consensus protocol to achieve agreement with the local estimates of all the other agents. To describe the \textit{optimize then agree} framework mathematically we define, the optimization scheme operator used by agent $i \in \mathbf{V}$ as the mapping $\mathcal{O}_i: \mathbb{R}^p \rightarrow \mathbb{R}^p$:
\begin{align}\label{eq:optoper}
    z^i(k) := \mathcal{O}_i(x^i(k-1)).
\end{align}
Similarly, define the consensus scheme as the operator mapping $\mathcal{C}: \mathbb{R}^{p\times n} \rightarrow \mathbb{R}^{p \times n}$:
\begin{align}\label{eq:consopt}
    \x(k) = \mathcal{C}(\z(k)),
\end{align}
where, $\z(k) = [z^1(k) \dots z^n(k)], \x(k) = [x^1(k) \dots x^n(k)] \in \mathbb{R}^{p \times n}$, such that $x^1(k) = x^2(k) = \dots = x^n(k)$.
Utilizing the above notation the \textit{optimize then agree} framework leads to the following algorithm:
\begin{algorithm}[h]
\footnotesize
    \SetKwBlock{Input}{Input:}{}
    \SetKwBlock{Initialize}{Initialize:}{}
    \SetKwBlock{STEPONE}{STEP 1:}{}
    \SetKwBlock{STEPTWO}{STEP 2:}{}
    \SetKwBlock{Repeat}{Repeat for $ k = 1,2, \dots$}{}
    \newcommand{\inparallel}{\textbf{In parallel}}
     \newcommand{\until}{\textbf{until}}
    \Repeat {
      \tcc{state updates using operators~(\ref{eq:optoper}) and~(\ref{eq:consopt})}
        $z^i(k) = \mathcal{O}_i(x^i(k-1)), \ \forall i\in \mathbf{V} $ \\
        $\x(k) = \mathcal{C}(\z(k))$
        }
      \until \ a stopping criterion is met
        \vspace{0.01in}
        \caption{Optimize then agree framework}
        \label{alg:optthenagree}
\end{algorithm}
}
\begin{remark}
We emphasize that the updates in Algorithm~\ref{alg:optthenagree} are general and do not depend on a particular choice of the optimization scheme or the consensus algorithm, thus the \textit{optimize then agree} framework provides a guideline for further exploration of distributed optimization algorithms. The framework of Algorithm~\ref{alg:optthenagree} can also be seen as an outer optimization loop and an inner consensus loop structure algorithm. Thus, existing schemes \cite{jakovetic2014fast, chen2012fast,johansson2008subgradient, berahas2018balancing} are a special case of the \textit{optimize then agree} framework with a particular choice of optimization scheme and a consensus protocol. 
\end{remark}
\textcolor{black}{
Choosing first-order gradient descent with a constant step-size $\alpha$ as the optimization scheme, Algorithm~\ref{alg:optthenagree} at any iteration $k + 1$ results in the following updates:
\begin{align}
    z^i(k+1) &= x^i(k) - \alpha \nabla f_i(x^i(k)), \ \forall i \in \mathbf{V},\nonumber \\
    \x(k+1) & = \mathcal{C}(\z(k+1)), \nonumber
\end{align}
with, the consensus operator implemented in a centralized manner (with the entire vector $\z(k+1)$ as the input). However, in order to make Algorithm~\ref{alg:optthenagree} applicable to distributed networks, we need to distributively realize~(\ref{eq:consopt}) of the consensus operator. To this end, one natural choice is to employ a distributed average consensus protocol \cite{xiao2004fast},\cite{kempe2003gossip}. Although these average consensus protocols lead to agreement among the agents to the initial average, the result holds only asymptotically which is a potential problem for an efficient implementation of Algorithm~\ref{alg:optthenagree}. To address this issue,
% we propose the gradient-consensus (\texttt{GradConsensus}) algorithm where
the consensus scheme operator is chosen to be a distributed finite-time approximate consensus protocol, which we call the $\varepsilon$-\texttt{Consensus} protocol \cite{khatana2020acc}. The $\varepsilon$-\texttt{Consensus} protocol produces a desired level of agreement among the estimates of all agents.  The accuracy in the agreement is determined by a parameter $\varepsilon>0$ independent of problem~(\ref{eq:introprob}) that can be chosen arbitrarily small. In the following section, we will first discuss the $\varepsilon$-\texttt{Consensus} protocol and then introduce the proposed \texttt{GradConsensus} algorithm.}

\begin{subsection}{$\varepsilon$-\texttt{Consensus} Protocol}\label{sec:epsilonCons}
The $\varepsilon$-\texttt{Consensus} protocol is a distributed finite-time terminated average consensus algorithm. The protocol was first proposed in earlier works \cite{prakash2019distributed}, \cite{saraswat2019distributed} by the authors in relation to consensus on scalar values. Here, we extend the protocol to higher dimensional average consensus protocols. Each agent $i \in \mathbf{V}$ has a vector $ z^i = [z_1^i \ z_2^i \dots z_p^i] \in \mathbb{R}^p$. The objective is to find an approximate estimate of the average $\overline{z} = \frac{1}{n}\sum_{i=1}^n z^i  \in \mathbb{R}^p$. To achieve this objective an iterative procedure is devised where each agent $j \in \mathbf{V}$ maintains a state variable $s^j(k) \in \mathbb{R}^p$ and an auxiliary variable $t_j(k) \in \mathbb{R}$, with the following initialization:
\begin{align}\label{eq:initialcond}
    s^j(0) = z^j, \ t_j(0) = 1, \ \text{for all} \ j \in \mathbf{V}.
\end{align}
\noindent Agent $j \in \mathbf{V}$ updates \cite{kempe2003gossip} 
according to: 
\begin{align}
  s^{j}(k+1)&=p_{jj}s^{j}(k)+\textstyle\sum_{\ell \in\mathit{N^{in}_{j}}}p_{j \ell }s^{\ell}(k)\label{eq:numerator}\\ 
  t_{j}(k+1)&=p_{jj}t_{j}(k)+\textstyle\sum_{\ell \in\mathit{N^{in}_{j}}}p_{j \ell }t_{\ell}(k)\label{eq:denominator}\\ 
  r^j(k+1) &= \textstyle\frac{1}{t_j(k+1)}s^j(k+1). \label{eq:ratio}
\end{align}
We make the following assumption on the graph $\mathbf{G}(\mathbf{V},\mathbf{E})$ and the associated matrix $\mathbf{P}=: [p_{ij}]$:  
\begin{assumption}\label{ass:graph_ass}
The directed graph $\mathbf{G}(\mathbf{V},\mathbf{E})$ is strongly-connected. The associated weighted adjacency matrix $\mathbf{P}=[p_{ij}]$ is a primitive, column-stochastic matrix. In particular, $ 0 \leq p_{ij} \leq 1$ and $\sum_{i=1}^n p_{ij} = 1$ for all $i,j \in \mathbf{V}$.
\end{assumption}
One choice of weights that satisfy Assumption~\ref{ass:graph_ass} is the Out-degree based equal neighbor weights rule \cite{olshevsky2009convergence}.
\noindent The convergence of the state $r^j(k):= \frac{1}{t_j(k)}s^j(k), j \in \mathbf{V}$ evolving according to~(\ref{eq:numerator})-(\ref{eq:ratio}), to the average $\overline{z} = \frac{1}{n}\sum_{i=1}^n z^i$ has been established rigorously  \cite{kempe2003gossip}, \cite{melbournepreprint}. We state the following result for updates~(\ref{eq:numerator})-(\ref{eq:ratio}):
\begin{theorem} \label{thm:consensusconv}
Let Assumption~\ref{ass:graph_ass} and the updates~(\ref{eq:numerator})-(\ref{eq:ratio}) hold. Let the initial conditions of the state and auxiliary variables $s^j$ and $t_j, j \in \mathbf{V}$ respectively be given by~(\ref{eq:initialcond}). Then 
$\lim_{k \rightarrow \infty} r^j(k) = \textstyle \frac{1}{n}\sum_{i=1}^n z^i, \ \text{for all} \ j \in \mathbf{V}.$
\end{theorem}
\noindent \textit{Proof.}
Refer \cite{melbournepreprint}, Theorem 2.1, for proof.\qed

\noindent We now provide a criteria for determining when consensus is reached within a tolerance $\varepsilon>0$. Let, $r^i(k)=\frac{1}{t_i(k)}s^i(k)$ and $\mathbf{r}(k) = [r^1(k) \ r^2(k) \dots r^n(k)] \in \mathbb{R}^{p \times n}$. At any iteration $k$, define the maximum $\overline{\mathbf{r}}(k)$ and minimum $\underline{\mathbf{r}}(k)$ state variable of the network over all the agents  as
\begin{align}
& \hspace{-0.2in} \overline{\mathbf{r}}(k) := \Big[ \underset{ 1 \leq j \leq n }{\max} \big\{\mathbf{r}_{[1j]}(k) \big\} \dots \ \underset{ 1 \leq j \leq n }{\max} \big\{\mathbf{r}_{[pj]}(k) \big\} \Big] \label{eq:maxval} \\
& \hspace{-0.2in} \underline{\mathbf{r}}(k) := \Big[ \underset{ 1 \leq j \leq n }{\min} \big\{\mathbf{r}_{[1j]}(k) \big\} \dots \ \underset{ 1 \leq j \leq n }{\min} \big\{\mathbf{r}_{[pj]}(k) \big\} \Big] \label{eq:minval},
\end{align}
where, $\mathbf{r}_{[ij]}(k)$ is $i^{th}$-row and $j^{th}$-column entry of $\mathbf{r}(k)$.

\begin{theorem}\label{thm:monotonic}
Let Assumption~\ref{ass:graph_ass} and updates~(\ref{eq:maxval}),~(\ref{eq:minval}) hold. Then, $\{\overline{\mathbf{r}}(u \mathbf{D})\}_{u \geq 1}$ and $\{\underline{\mathbf{r}}(u \mathbf{D})\}_{u \geq 1}$ are strictly monotonic sequences converging to $\overline{z} = \frac{1}{n}\sum_{i=1}^n z^i$.
\end{theorem}
\textbf{Proof.}
Refer \cite{melbournepreprint}, Theorem 4.1, for Proof. \qed

\noindent To distributively determine the sequences of global maximum $\{\overline{\mathbf{r}}(u \mathbf{D})\}_{u \geq 1}$ and minimum $\{\underline{\mathbf{r}}(u \mathbf{D})\}_{u \geq 1}$ each agent $j \in \mathbf{V}$, maintains two variables $M^j(k), m^j(k) \in \mathbb{R}^p$ at any iteration $k$. The variables $M^j$ and $m^j$ are initialized as $M^j(0)  = m^j(0) = z^j \ \text{for all} \ j \in \mathbf{V}$ and are updated as:
\begin{align}
    & \hspace{-0.12in} M^j(k+1)  = \Big[ \underset{ \ell \in N_j^{in} \cup j }{\max} M^\ell_1(k) \dots  \underset{ \ell \in N_j^{in} \cup j }{\max} M^\ell_p(k) \Big] \label{eq:maxupdate} \\
    & \hspace{-0.1in} m^j(k+1)  = \Big[ \underset{ \ell \in N_j^{in} \cup j}{\max} m^\ell_1(k) \dots  \underset{ \ell \in N_j^{in} \cup j }{\max} m^\ell_p(k) \Big], \label{eq:minupdate}
\end{align}
where, $M^\ell_i(k)$, and $m^\ell_i(k)$ are the $i^{th}$ entry of $M^\ell(k)$ and $m^\ell(k)$ respectively. 

\begin{proposition}\label{prop:mxp-mnp}
Let Assumption~\ref{ass:graph_ass}, updates~(\ref{eq:maxupdate}) and~(\ref{eq:minupdate}) hold. Then the variables $M^j(k)$ and $m^j(k)$  for all $j \in \mathbf{V}$, converges to $\overline{\mathbf{r}}(0)$ and $\underline{\mathbf{r}}(0)$ respectively in finite number of iterations $k_f \leq \mathbf{D}$.
\end{proposition}
\noindent \textit{Proof.}
Refer \cite{yadav2007distributed}, Proposition 2.3, for proof. \qed

\textcolor{black}{
Proposition~\ref{prop:mxp-mnp} leads to a method for finding the sequences $\{\overline{\mathbf{r}}(u \mathbf{D})\}_{u \geq 1}$ and $\{\underline{\mathbf{r}}(u \mathbf{D})\}_{u \geq 1}$ distributively: re-initializing the updates~(\ref{eq:maxupdate}),~(\ref{eq:minupdate}) for all agents $j \in \mathbf{V}$, at every iteration of the form $k = u\mathbf{D}, u = 1, 2, \dots$, to the variable $r^j(u\mathbf{D})$ will allow the variables $M^j$ and $m^j$ to converge to $\overline{\mathbf{r}}((u-1)\mathbf{D})$ and $\underline{\mathbf{r}}((u-1)\mathbf{D})$ respectively, after a finite number of iterations $k_f \leq \mathbf{D}$ by Proposition~\ref{prop:mxp-mnp}. Since, both the sequences $\{\overline{\mathbf{r}}(m \mathbf{D})\}_{m \geq 1}$ and $\{\underline{\mathbf{r}}(m \mathbf{D})\}_{m \geq 1}$ converge to $\overline{z}$ (Theorem~\ref{thm:monotonic}), the norm of the difference between the two, i.e., $\|\overline{\mathbf{r}}(m \mathbf{D}) - \underline{\mathbf{r}}(m \mathbf{D})\|$ also converges to zero. Therefore, given $\varepsilon > 0$, the quantity $\|\overline{\mathbf{r}}(m \mathbf{D}) - \underline{\mathbf{r}}(m \mathbf{D})\|$ will eventually become less than $\varepsilon$. Based on the above observations we propose an algorithm which allows all agents to simultaneously converge to an $\varepsilon$-close estimate of $\overline{z}$ with $\varepsilon$ being an arbitrary pre-specified tolerance. For $u = 1,2, \dots$, let $\gamma^j(u \mathbf{D}) := M^j(u \mathbf{D}) - m^j(u \mathbf{D})$. 
Each agent computes the norm $\| \gamma^j(u \mathbf{D}) \|$ every $\mathbf{D}$ iterations and compares it with $\varepsilon$. If $\| \gamma^j(\tilde{u} \mathbf{D})\| := \|M^j(\tilde{u} \mathbf{D}) - m^j(\tilde{u} \mathbf{D})\| < \varepsilon$ at any iteration $\tilde{u} \mathbf{D}$, then agent $j$, knowing its estimate $r^j(\tilde{u} \mathbf{D})$ is $\varepsilon$-close to $\overline{z}$, terminates the consensus protocol updates~(\ref{eq:numerator})-(\ref{eq:ratio}) at $\tilde{u} \mathbf{D}$. The variable $r^j(\tilde{u} \mathbf{D})$ is the $\varepsilon$-close estimate of $\overline{z}$ available with agent $j$. 
}
% \begin{remark}\label{rem:globalparameter}
% Notice the only global parameter needed for Algorithm~\ref{alg:epsiloncons} is an upper bound $D$ on the diameter of the directed graph $\mathbf{G}(\mathbf{V},\mathbf{E})$, which is a design parameter and thus readily available in most applications. 
% \end{remark}

\end{subsection}

\begin{subsection}{The \texttt{GradConsensus} Algorithm}\label{sec:GCalgorithm}
Here, each agent $i$ maintains two estimates: an optimization variable $x^i(k)\in \mathbb{R}^p$ and a gradient-descent update variable $z^i(k) \in \mathbb{R}^p$ at any iteration $k$. Every iteration $k$ proceeds in two steps: first every agent $i$ updates $z^i(k)$ via a gradient descent update based on its own function $f_i$ at $x^i(k-1)$. At the next step, the optimization variable $x^i(k)$ is updated to an estimate which is $\varepsilon(k)$-close to the average value $\hat{z}(k):= \frac{1}{n} \sum_{i=1}^n z^i(k)$, i.e. $\|x^i(k) - \hat{z}(k)\| \leq \varepsilon(k)$, using the distributed $\varepsilon$-\texttt{Consensus} protocol, initialized with $z^i(k)$ as the initial condition for the agent $i$ and tolerance $\varepsilon(k)$.
\begin{algorithm}[h]
% \scriptsize
    \SetKwBlock{Input}{Input:}{}
    \SetKwBlock{Initialize}{Initialize:}{}
    \SetKwBlock{STEPONE}{STEP 1:}{}
    \SetKwBlock{STEPTWO}{STEP 2:}{}
    \SetKwBlock{Repeat}{Repeat for $ k = 1,2, \dots$}{}
    \newcommand{\inparallel}{\textbf{In parallel}}
    \newcommand{\until}{\textbf{until}}
    \Input{ Choose consensus tolerances $\{\varepsilon(k)\}_{k \geq 0}$ and step-size $\alpha$}
    \Initialize{ 
        - Each agent $j \in \mathbf{V}$ initializes its estimates as $x^j(0) = z^j(0) = 0.$
      }
    \Repeat {
    \For{$j = 1,2,3,\dots,n $, (\inparallel)}{
        {\fontfamily{pcr}\selectfont \small{outer gradient descent iteration:}}
        $z^i(k) = x^i(k-1) - \alpha \nabla f_i(x^i(k-1))$ 
        {\fontfamily{pcr}\selectfont \small{inner consensus iterations:}}
        $x^i(k) \leftarrow \varepsilon(k)$-\texttt{Consensus}$(z^i(k), \ i \in \mathbf{V})$
        }
        }
        \until \ a stopping criterion is met
        \vspace{0.01in}
        \caption{\texttt{GradConsensus} Algorithm}
        \label{alg:gradcons}
\end{algorithm}
\textcolor{black}{Thus, at each iteration $k$ of Algorithm~\ref{alg:gradcons}, every individual agent $ i \in \mathbf{V}$ performs the following updates:
\begin{align}
    z^i(k) &= x^i(k-1)  - \alpha \nabla f_i(x^i(k-1)) \label{eq:gradStep}\\
    x^i(k) &= r^i(k_c(k)), \mbox{with,} \\
    \| r^i(k_c(k)) &- \hat{z}(k) \| < \varepsilon(k), \textstyle \hat{z}(k):= \frac{1}{n} \sum_{i=1}^n z^i(k), \label{eq:conStep}
\end{align}
where, $k_c(k)$ denotes the number of iterations required by the $\varepsilon$-\texttt{Consensus} protocol to reach the consensus accuracy within $\varepsilon(k)$ at iteration $k$ of Algorithm~\ref{alg:gradcons}, and $r^i(k_c(k))$ is the approximate estimate of $\hat{z}(k)$ produced by the $\varepsilon$-\texttt{Consensus} protocol.} 
\begin{remark}
To keep the analysis concise we assume the step-size $\alpha$ to be the same for all the agents. However, this does not pose any restriction to scheme. Before the start of Algorithm~\ref{alg:gradcons} any agent can choose a step-size $\alpha$ and by using the Maximum Consensus Protocol \cite{prakash2018distributed} (one dimensional equivalent of update~(\ref{eq:maxupdate})) each agent can know the value of step-size $\alpha$ within $\mathbf{D}$ number of iterations (see Proposition~\ref{prop:mxp-mnp}). After the step-size is known each agent can execute Algorithm~\ref{alg:gradcons} independently. 
\end{remark}
\begin{remark}
Note that \texttt{GradConsensus} algorithm (updates~(\ref{eq:numerator})-(\ref{eq:ratio})) utilize a column stochastic matrix which allows for a distributed synthesis of the protocol. This feature gives an advantage over existing schemes in the literature \cite{nedic2009distributed,shi2015extra,jakovetic2014fast,qu2019accelerated} that utilize a doubly-stochastic weight matrix and cannot achieve tractable distributed synthesis in directed networks.  
\end{remark}
\end{subsection}

\end{section}

\begin{section}{Convergence Analysis for \texttt{GradConsensus}}\label{sec:convgAnalysis}
This section is dedicated to the analysis of the proposed \texttt{GradConsensus} algorithm. \textcolor{black}{We will analyze the convergence of the \texttt{GradConsensus} algorithm for two scenarios: (i) when the aggregate function $f$ is convex and Lipschitz differentiable and, (ii) when $f$ is strongly convex and Lipschitz differentiable. In both the scenarios we establish the convergence of the iterates generated by Algorithm~\ref{alg:gradcons} to the optimal solution of problem~(\ref{eq:introprob}). We will provide estimates of the rate of convergence to the optimal solution in terms of the (outer) iterations $k$ of the Algorithm~\ref{alg:gradcons} and the total communication steps $\mathcal{K}$.} We begin presenting preliminary results that we will utilize in the convergence analysis. 
\begin{subsection}{Supporting Lemmas}
We make the following assumptions throughout the rest of the article:
\begin{assumption}\label{ass:finite_fstar} For problem~(\ref{eq:introprob}) the optimal value $f^*$ is finite and the optimal solution set $\mathcal{X}^*$ is non-empty. 
\end{assumption}
\begin{assumption}\label{ass:fi_conv_lip}
\begin{enumerate}
    \item For all $\ i \in \mathbf{V}$, $f_i$ is a proper closed-convex function with a lower bound. 
    \item Each function $f_i$ is Lipschitz differentiable with constant $L_{f_i} > 0$. 
\end{enumerate}
\end{assumption}
\textcolor{black}{
We make the following assumption which primarily effects the analysis of number of inner consensus iterations needed by the Algorithm~\ref{alg:gradcons}. 
\begin{assumption}\label{ass:bounded gradient}
\item The gradients of functions $f_i$ are bounded, i.e., there exists $h_i < \infty$ such that $\forall x \in \mathbb{R}^p, \|\nabla f_i(x)\| \leq h_i$.
\end{assumption}We establish most results with Assumption~\ref{ass:bounded gradient}, however, the convergence analysis and rate estimates of the \texttt{GradConsensus} remains valid without Assumption~\ref{ass:bounded gradient}. Relaxing Assumption~\ref{ass:bounded gradient} only effects the number of iterations required by the $\varepsilon(k)$-\texttt{Consensus} protocol. We state results that hold in the absence of Assumption~\ref{ass:bounded gradient} in remarks after each Theorem.
}
Let the average of the optimization variables at iteration $k \geq 0$ be denoted as: $\hat{x}(k) := \textstyle \frac{1}{n}\sum^{n}_{i=1} x^i(k).$
% \begin{align}\label{eq:xhat}
%   \hat{x}(k) := \textstyle \frac{1}{n}\sum^{n}_{i=1} x^i(k).
% \end{align}
We denote the gradient of the function $f$ evaluated at the individual optimization variables of all the $n$ agents and at the average $\hat{x}(k)$ at any iteration $k \geq 0$ as:
\begin{align*}
  g(k) := \textstyle \sum^{n}_{i=1}\nabla f_i(x^i(k)), \  \ \hat{g}(k) := \textstyle \sum^{n}_{i=1}\nabla f_i(\hat{x}(k)).
\end{align*}
Each iteration of Algorithm~\ref{alg:gradcons} utilizes an $\varepsilon$-\texttt{Consensus} protocol. The following Lemma provides the number of communication steps required by the $\varepsilon$-\texttt{Consensus} protocol to converge to an $\varepsilon(k)$-close solution at the $k^{th}$ outer gradient descent iteration of Algorithm~\ref{alg:gradcons}.
\textcolor{black}{
\begin{lemma}\label{lem:cons_comm}
Let Assumptions~\ref{ass:graph_ass},~\ref{ass:fi_conv_lip} and~\ref{ass:bounded gradient} hold. Then at any outer gradient descent iteration $k$ of Algorithm~\ref{alg:gradcons}, after $k_c(k) = \ceil*{  \left[\frac{\log\big(\frac{1}{\varepsilon(k)}\big) }{-\log \lambda}+ \frac{\log\big(\frac{8n}{\delta}(\sum_{s=0}^{k-1} \varepsilon(s) + \alpha k h_m) \big)}{-\log \lambda}\right] }$ iterations of the consensus protocol (updates~(\ref{eq:numerator})-(\ref{eq:ratio})) with the initial condition $s^i(0) = z^i(k), \ t_i(0) = 1, \forall i \in \mathbf{V}$  we have: 
\begin{align*}
 \|x^i(k) - \hat{z}(k)\|  \leq \textstyle \varepsilon(k), \ \forall i \in \mathbf{V},
\end{align*}
where, $\varepsilon(k)$ is the consensus tolerance parameter, $h_m := \max_{1 \leq i\leq n} h_i$ and $\delta >0$, $\lambda \in (0,1)$ are parameters of the graph $\mathbf{G}$ satisfying $
    \delta \geq \frac{1}{n^n}, \ \ \lambda \leq \textstyle \left(1 - \frac{1}{n^n} \right)$.
\end{lemma}
\textbf{Proof.}
To begin, we will present a modification of an existing result (Lemma 1 \cite{nedic2014distributed}). The result in Lemma 1 \cite{nedic2014distributed}, with the perturbation term being zero, reduces to a convergence result for the push-sum protocol. Using this property, we conclude that the updates~(\ref{eq:numerator})-(\ref{eq:ratio}) converges at a geometric rate to the average of the initial values. Note that the \texttt{GradConsensus} algorithm at every iteration $k$ utilizes $\varepsilon$-\texttt{Consensus} protocol (updates~(\ref{eq:numerator})-(\ref{eq:ratio})) with the initial values $z^i(k)$. Hence, the estimates $x^i(k)$ converges to the average $\hat{z}(k)$ (of the initial values) at a geometric rate. Therefore, we conclude,
\begin{align}\label{eq:nedic_result}
    \hspace{-0.098in} \|x^i(k) - \hat{z}(k)\| \leq \textstyle \frac{8 \sqrt{n} \lambda^{k_c(k)}}{\delta} \|\z(k)\|, \ \forall i \in \mathbf{V},
\end{align} 
where, $\z (k) := [z^1(k), \dots, z^n(k)] \in \mathbb{R}^{p\times n}$ and, $\delta >0, \lambda \in (0,1)$ satisfy: $
    \delta \geq \frac{1}{n^n}, \ \ \lambda \leq \textstyle \left(1 - \frac{1}{n^n} \right)$.
Here, the variables $\lambda$ and $\delta$ are parameters of the graph $\mathbf{G}$. The parameter $\lambda$ measures the speed at which the graph $\mathbf{G}$ 
diffuses the information among the agents over time. For a regular graph $\mathbf{G}$ (leading to a symmetric doubly stochastic $\mathbf{P}$) $\lambda$ is equivalent to the second largest eigenvalue of $\mathbf{P}$. Further, the parameter, $\delta$ measures the imbalance of influences among the in $\mathbf{G}$ \cite{nedic2014distributed}. Next, we will bound $\|\z(k)\|$ using an induction argument.\\
\textit{Claim:} Under Assumption~\ref{ass:bounded gradient}, at any $k$, 
$\|z^i(k)\| \leq \sum_{s=1}^{k-1} \varepsilon(s) + \alpha k h_m$, for all $i \in \mathbf{V}$.\\ \textit{Proof:} For $k = 1$, for any $i$ we have,  $\| z^i(1) \| = \| x^i(0) - \alpha \nabla f_i(x^i(0))\| \leq \alpha \max_{1 \leq i \leq n} h_i = \alpha h_m$. Assume, for $k = k, \| z^i(k) \| \leq \sum_{s=1}^{k-1} \varepsilon(s) + \alpha k h_m$, for all $i \in \mathbf{V}$. Now, for $k = k+1$, for any $i$, $\| z^i(k+1) \| = \| x^i(k) - \alpha \nabla f_i(x^i(k))\| \leq \| x^i(k) - \hat{z}(k)\| + \| \hat{z}(k)\| + \alpha \| \nabla f_i(x^i(k))\| \leq \varepsilon(k) + \sum_{s=1}^{k-1} \varepsilon(s) + \alpha k h_m + \alpha \max_{1 \leq i \leq n} h_i = \sum_{s=1}^{k} \varepsilon(s) + \alpha (k+1) h_m$, for all $i \in \mathbf{V}$, where we used~(\ref{eq:conStep}) in the last inequality. Therefore, induction holds. \\
Using the above claim it can be shown that $\|\z(k)\| \leq \sqrt{n}(\sum_{s=0}^{k-1} \varepsilon(s) + \alpha k h_m)$. 
If $\frac{\varepsilon(k) \delta}{\lambda^k_c(k) 8 \sqrt{n}} = \|\z(k)\| \leq \sqrt{n}(\sum_{s=0}^{k-1} \varepsilon(s) + \alpha k h_m)$  it implies that, $\frac{\delta \varepsilon(k)}{8n \sum_{s=0}^{k-1} \varepsilon(s) + \alpha k h_m}$ $\leq \lambda^{k_c(k)}$. Therefore, we have,
\begin{align*}
    k_c(k) &\leq \textstyle \frac{-1}{\log \lambda} \Big[\log\big(\frac{1}{\varepsilon(k)}\big) \nonumber \\
    & \hspace{0.4in} \textstyle + \log\big(\frac{8n}{\delta}(\sum_{s=0}^{k-1} \varepsilon(s) + \alpha k h_m) \big)\Big]:= \overline{k_c}(k).
\end{align*}
Using~(\ref{eq:nedic_result}) we conclude that after $\overline{k_c}(k)$ number of iterations at the $k^{th}$ outer gradient descent iteration, $\|x^i(k) - \hat{z}(k)\| \leq \varepsilon(k)$, for all $i \in \mathbf{V}$.
\begin{remark}
Lemma~\ref{lem:cons_comm} provides an upper bound on the number of communication steps required at the $k^{th}$ outer gradient descent iteration of Algorithm~\ref{alg:gradcons} to obtain $\varepsilon(k)$-close solution. In particular, if $\sum_{k=0}^{\infty} \varepsilon(k) < \infty$, then at the $k^{th}$ outer gradient descent iteration of Algorithm~\ref{alg:gradcons}, after $O\big(\log\big(\frac{1}{\varepsilon(k)}\big) + \log(k)\big)$ communication steps the estimates of all the agents are guaranteed to be $\varepsilon(k)$-close to each other.  
\end{remark}
\begin{remark}
Note that the result in Lemma~\ref{lem:cons_comm} makes use of Assumption~\ref{ass:bounded gradient}. However, this restriction is not present for a wide variety of scenarios; here an upper bound on the gradient of the functions can be obtained if the gradient descent minimization step is performed over a compact set $\mathcal{X}$ with a diameter $R$. In particular, since, $f_i$ have Lipschitz continuous gradients, 
\begin{align*}
    & \|\nabla f_i(x^i(k)) - \nabla f_i(x^i(0))\| \leq \max_{1 \leq i \lq n} L_{f_i}\|x^i(k) - x^i(0)\| \\
    & \implies \|f_i(x^i(k))\| \leq L_h R + \max_{1 \leq i \leq n} \|\nabla f_i(x^i(0))\| := h_m , 
\end{align*}
where, $R:= \sup_{x, y \in \mathcal{X}} \|x-y\|$ is the diameter of the set $\mathcal{X}$. The $\varepsilon$-\texttt{Consensus} protocol can be utilized to get an $\varepsilon(k)$-close solution within finite number of iterations without Assumption~\ref{ass:bounded gradient}. This is established in Lemma~{6.1} presented in Appendix where the $O(k\log k)$ is replaced by $O(k^2)$ for the number of communication iterates.
\end{remark}
A consequence of the $\varepsilon$-\texttt{Consensus} protocol is that the difference between $g(k)$ and $\hat{g}(k)$ is bounded for sufficiently large consensus loop iterations $k_c(k)$. The next Lemma establishes this property of the \texttt{GradConsensus} algorithm.
\begin{lemma}\label{lem:gradbound} 
Under assumptions~\ref{ass:graph_ass} and~\ref{ass:fi_conv_lip}, at $k^{th}$ outer gradient descent iteration of Algorithm~\ref{alg:gradcons} after $\overline{k}_c(k)$ iterations of the consensus protocol, we have: 
\begin{align*}
 \|g(k) - \hat{g}(k)\|  \leq \textstyle 2n L_h \varepsilon(k),
\end{align*}
where, $L_h = \max_{1\leq i \leq n} L_{f_i}, \varepsilon(k)$ is the consensus tolerance parameter, and $\overline{k}_c(k)$ is as defined in Lemma~\ref{lem:cons_comm}.
\end{lemma}
\textbf{Proof.}
\textcolor{black}{
Under Assumption~\ref{ass:fi_conv_lip}, we note that,
\begin{align}
    \|g(k) - \hat{g}(k)\| &= \textstyle \left\|\sum_{i=1}^{n}\nabla f_i(x^i(k)) - \sum_{i=1}^{n}\nabla f_i(\hat{x}(k))\right\| \nonumber \\
    & \textstyle \leq L_h \sum_{i=1}^{n} \|x^i(k) - \hat{x}(k) + \hat{z}(k) - \hat{z}(k)\| \nonumber \\
    & \hspace{-0.15in} \textstyle \leq L_h \sum_{i=1}^{n} \|x^i(k) - \hat{z}(k) \| + \|\hat{z}(k) - \hat{x}(k)\| \nonumber \\
    & \leq 2n L_h \varepsilon(k) \label{eq:intermediate}, 
\end{align}
where, in the last step follows from Lemma~\ref{lem:cons_comm}.
\qed
}
}

\textcolor{black}{
From update~(\ref{eq:gradStep}), 
\begin{align}
   \textstyle \frac{1}{n} \sum_{i=1}^n z^i(k) & =  \textstyle  \frac{1}{n} \sum_{i=1}^n \left[x^i(k-1)  - \alpha \nabla f_i(x^i(k-1))\right] \nonumber \\
   & = \hat{x}(k-1) - \textstyle \frac{\alpha}{n} \sum_{i=1}^n \nabla  f_i(x^i(k-1)).\nonumber 
\end{align}
\textcolor{black}{Therefore, we have
\begin{align}
   \hat{z}(k)  &= \hat{x}(k-1) - \textstyle \hat{\alpha}g(k-1) + \hat{x}(k) - \hat{x}(k), \implies \nonumber\\
   \hat{x}(k) &= \hat{x}(k-1) - \textstyle \hat{\alpha}g(k-1) + v(k), \label{eq:closedform_upd}
\end{align}
where, $ v(k):= \hat{x}(k) - \hat{z}(k), \|v(k)\| \leq \varepsilon(k)$ due to Lemma~\ref{lem:cons_comm} and $\hat{\alpha} = \frac{\alpha}{n}$.
}
In the centralized setting the information about the gradient of the function $f$, i.e. $\nabla f(x) = \sum_{i=1}^n \nabla f_i(x) = \hat{g}(x)$ is known to the central server. Here, an iteration of the (centralized) gradient descent will be of the form: $\tilde{x} = x - \alpha \hat{g}(x)$, where, $\tilde{x}$ denotes the updated estimate of the optimal solution. Note that due to Lemma~\ref{lem:gradbound} the update~(\ref{eq:closedform_upd}) can be viewed as an inexact centralized gradient descent update performed at the average of all the agents' estimates for the function $f$. In particular,  
\textcolor{black}{
\begin{align*}
    \hat{x}(k+1) &= \hat{x}(k) - \textstyle \hat{\alpha}\hat{g}(k) + u(k), \ \mbox{with}\\
     u(k):= \hat{\alpha} (\hat{g}(k) &- g(k)) + v(k),  \|u(k)\| \leq (2L_h \alpha + 1) \varepsilon(k).
\end{align*} 
}
Therefore, the \texttt{GradConsensus} algorithm performs an approximate centralized gradient descent updates at each iteration. Due to this property, Algorithm~\ref{alg:gradcons} exhibits convergence properties similar to a centralized gradient descent method. We define the solution residual $\hat{e}(k)$ and objective value residual $\hat{r}(k)$ at iteration $k$ as:
\begin{align}
    \hat{e}(k) &:= \hat{x}(k) - x^*, \label{eq:ehat} \ \mbox{and} \\
    \hat{r}(k) &:= f(\hat{x}(k)) - f^* = \textstyle \sum^{n}_{i=1}f_i(\hat{x}(k)) - f^*\label{eq:rhat}.
\end{align}
Under Assumption~\ref{ass:fi_conv_lip} since all $f_i$ are Lipschitz differentiable with parameter $L_{f_i}$, $f$ also is Lipschitz differentiable with the constant $L_{f} :=  \sum_{i=1}^n L_{f_i}$.
}
The following two Lemmas are properties of Lipschitz differentiable convex and strongly convex functions that are standard results in the convex analysis. We will make use of these identities in Theorems~\ref{thm:conproof} and~\ref{thm:strconvproof}.
\begin{lemma}\label{lem:conv_inner_prod_bound}
Under Assumption~\ref{ass:fi_conv_lip} for all $x, y \in \mathbb{R}^p$, $ \\ \langle \nabla f(y) - \nabla f(x), y - x \rangle  \geq \textstyle\frac{1}{L_f}\|\nabla f(y) - \nabla f(x)\|^2.$
% \begin{align*}
%     \langle \nabla f(y) - \nabla f(x), y - x \rangle &  \geq \textstyle\frac{1}{L_f}\|\nabla f(y) - \nabla f(x)\|^2, 
% \end{align*}
% where, $L_f = \sum_{i=1}^n L_{f_i}$.
\end{lemma}
\textbf{Proof.} Refer \cite{nesterov2013introductory}, Theorem 2.1.5, for proof. \qed

\begin{lemma}\label{lem:strconv_inner_prod_bound}
Under Assumptions~\ref{ass:fi_conv_lip}, ~\ref{ass:f_strconv} for all $x, y \in \mathbb{R}^p$,\\
$\langle x - y, \nabla f(x) - \nabla f(y) \rangle \geq \textstyle \frac{1}{\sigma+L_{f}} \|\nabla f(x) - \nabla f(y)\|^2 + \textstyle \frac{\sigma L_{f}}{\sigma  +L_{f}}\|x-y\|^2.$
%     \begin{align*}
%     \langle x - y, \nabla f(x) - \nabla f(y) \rangle &  \geq \textstyle \frac{1}{\sigma+L_{f}} \|\nabla f(x) - \nabla f(y)\|^2 \\
%       & \hspace{0.65in} + \textstyle \frac{\sigma L_{f}}{\sigma  +L_{f}}\|x-y\|^2.
% \end{align*}
\end{lemma}
\textbf{Proof.} Refer \cite{nesterov2013introductory}, Theorem 2.1.12, for proof. \qed
\end{subsection}

\begin{subsection}{Convergence Analysis for Convex $f$}
\textcolor{black}{\textcolor{black}{In this subsection, we present the convergence result for the \texttt{GradConsensus} when the function $f$ is convex and Lipschitz differentiable. Under these assumptions Theorem~\ref{thm:conproof} establishes a Q-linear rate of convergence to an $O(\varepsilon)$ neighborhood of the optimal solution. Moreover, we also provide a bound on the total number of communication steps required by the \texttt{GradConsensus} algorithm to achieve the convergence rate estimates.}
\textcolor{black}{
\begin{theorem}\label{thm:conproof}
Let assumptions~\ref{ass:graph_ass}-\ref{ass:bounded gradient} hold. Let $\hat{\alpha} \leq \frac{2}{L_f}$ and $\varepsilon(k) = \frac{\varepsilon_0}{k^{1+\eta}}$, where, $\varepsilon_0, \eta \in (0,1)$ are positive constants. Consider, the outer gradient descent iteration $k$, the total consensus communication iterations $\mathcal{K}:= \sum_{s=1}^k k_c(s)$ are bounded by $\sum_{s=0}^k \left[ \frac{\log\big(\frac{s^{1+\eta}}{\varepsilon_0}\big)}{-\log \lambda} + \frac{\log\big(\frac{8n}{\delta}( \varepsilon_0 \zeta(1+\eta) + \alpha s h_m) \big)}{-\log \lambda}\right] = 
O(k\log k)$, where, $\zeta(.)$ denotes the Riemann zeta function. At the $k^{th}$ outer gradient descent iteration of Algorithm~\ref{alg:gradcons}, if the objective function residual (defined in~(\ref{eq:rhat})), $\hat{r}(k)> 2\mathbf{e} \sqrt{\frac{4\alpha n L_h^2 + L_f +2/\hat{\alpha}}{\hat{\alpha}}} \varepsilon_0$, where, $ \mathbf{e} := \|\hat{x}(0) - x^*\| + \textstyle (2\alpha \varepsilon_0 L_h +1 )\zeta(1+\eta)$, then $\hat{r}(k)$ decreases at a Q-linear rate, with respect to the outer gradient descent iterations. In particular, there exists $\beta \in (0,1)$, such that at the $k^{th}$ outer gradient descent iteration $\hat{r}(k) \leq \beta^k \hat{r}(0)$. 
\end{theorem}
\textbf{Proof.} We start by showing that the solution residual $\|\hat{e}(k)\|$ is bounded. Consider,
\begin{align}
    \|\hat{x}(k) - \hat{\alpha}\hat{g}(k) - x^*\|^2 & = \   \|\hat{x}(k) - x^*\|^2  + \hat{\alpha}^2 \|\hat{g}(k)\|^2 \nonumber \\
    & \hspace{0.56in} - 2\hat{\alpha} \langle \hat{g}(k), \hat{x}(k) - x^* \rangle\nonumber \\
    & \hspace{-1.22in} \leq \textstyle \|\hat{x}(k) - x^*\|^2 + \textstyle  \hat{\alpha}\left(\hat{\alpha} - \frac{2}{ L_f} \right)\|\hat{g}(k)\|^2 \nonumber\\
    & \hspace{-1.22in} \leq \|\hat{x}(k) - x^*\|^2\label{eq:dec_e_hat}, 
\end{align}
where, we used Lemma~\ref{lem:conv_inner_prod_bound} and the fact $\hat{g}(x^*) = 0.$ Using~(\ref{eq:closedform_upd}),  Lemma~\ref{lem:gradbound} and~(\ref{eq:dec_e_hat}), 
% Since, $k_c(k) \geq \frac{\log k^{2+\zeta}}{-\log (\lambda)} + \frac{\log (\frac{8nh_m}{\delta \varepsilon})}{-\log (\lambda)}$, then using Lemma~\ref{lem:gradbound} and~(\ref{eq:dec_e_hat}), 
\begin{align} 
    \|\hat{x}(k+1) - x^*\| & = \|\hat{x}(k) - \hat{\alpha} g(k) - x^* + v(k)\| \nonumber \\
    & \hspace{-0.4in} =  \|\hat{x}(k) - \hat{\alpha} g(k) + \hat{\alpha} \hat{g}(k) - \hat{\alpha} \hat{g}(k) - x^* + v(k)\| \nonumber \\
    & \hspace{-0.45in}  \leq \|\hat{x}(k) - \hat{\alpha} \hat{g}(k) - x^*\| + \hat{\alpha}\| g(k) - \hat{g}(k) \| + \varepsilon(k) \nonumber \\
    &\leq \|\hat{x}(k) - x^*\| + \textstyle \frac{2\alpha \varepsilon_0 L_h + 1}{k^{1 +\eta}} \nonumber \\
    & \hspace{-0.2in} \leq \|\hat{x}(0) - x^*\| + (2\alpha \varepsilon_0 L_h + 1) \textstyle \sum_{s=0}^k \frac{1}{k^{1+\eta}} \nonumber \\
    & \hspace{-0.7in} \leq \|\hat{x}(0) - x^*\| + \textstyle (2\alpha \varepsilon_0 L_h + 1) \zeta(1+\eta) := \mathbf{e},\label{eq:ehat_bound}\\
    &\hspace{-1in}\mbox{where, $\zeta(.)$ is the Riemann zeta function. Further,} \nonumber\\
    f(\hat{x}(k+1)) &\leq f(\hat{x}(k)) + \langle \hat{g}(k), \hat{x}(k+1)-\hat{x}(k)\rangle \nonumber \\ 
    & \hspace{0.65in} + \textstyle \frac{L_f}{2}\|\hat{x}(k+1) - \hat{x}(k)\|^2 \nonumber \\
   \implies \hat{r}(k+1)  &\leq  \hat{r}(k) + \langle \hat{g}(k), \hat{x}(k+1)-\hat{x}(k)\rangle \nonumber \\
    & \hspace{0.65in} + \textstyle \frac{L_f}{2}\|\hat{x}(k+1) - \hat{x}(k)\|^2 \nonumber\\
    & \hspace{-0.4in} = \hat{r}(k) - \hat{\alpha} \langle \hat{g}(k), g(k)\rangle + \textstyle \frac{\hat{\alpha}^2L_f}{2}\|g(k)\|^2 \nonumber \\
    & \textstyle  + (1-\hat{\alpha}L_f) \langle g(k), v(k) \rangle  + \frac{L_f}{2} \|v(k)\|^2 \nonumber\\
    & \hspace{-0.65in} = \hat{r}(k) - \hat{\alpha} \langle \hat{g}(k), \hat{g}(k)\rangle + \textstyle \frac{\hat{\alpha}^2L_f}{2} \| \hat{g}(k) - g(k)\|^2 \nonumber \\
    & \hspace{-0.58in} + \textstyle \frac{\hat{\alpha}^2L_f}{2} \|\hat{g}(k)\|^2 + \textstyle ( \hat{\alpha} - \hat{\alpha}^2 L_f ) \langle  \hat{g}(k), \hat{g}(k) - g(k) \rangle \nonumber \\ 
    & \textstyle  + (1-\hat{\alpha}L_f) \langle g(k), v(k) \rangle  + \frac{L_f}{2} \varepsilon(k)^2 \nonumber \\
    & \hspace{-0.65in} \leq \hat{r}(k) - \textstyle \frac{\hat{\alpha}}{2}\| \hat{g}(k) \|^2 + \frac{4\alpha n L_h^2 + L_f }{2} \varepsilon(k)^2, \nonumber \\
    & \textstyle  + (1-\hat{\alpha}L_f) \langle g(k), v(k) \rangle \nonumber \\
    & \hspace{-0.65in} \leq \hat{r}(k) - \textstyle  \frac{\hat{\alpha}}{4} \| \hat{g}(k) \|^2+  \frac{4\alpha n L_h^2 + L_f +2/\hat{\alpha}}{2} \varepsilon(k)^2
\end{align}    
where, we used the Cauchy-Schwarz and AM-GM inequality to get the second last inequality and inequality $\pm 2\langle a, b \rangle \leq  \zeta\|a\|^2 + \frac{1}{\zeta}\|b\|^2,$ $a,b \in \mathbb{R}^p$ with $\zeta = \frac{2}{\hat{\alpha}}$ to get the last inequality. Note that, $\hat{r}(k) = f(\hat{x}(k)) - f^* \leq \langle \hat{g}(k), \hat{x}(k) - x^* \rangle = \langle \hat{g}(k), \hat{e}(k) \rangle$. Using~(\ref{eq:ehat_bound}) we obtain,
\begin{align*}
    \|\hat{g}(k)\| \geq \|\hat{g}(k)\| \textstyle \frac{\|\hat{x}(k) - x^*\|}{\mathbf{e}} \geq \frac{\langle \hat{g}(k), \hat{x}(k) - x^* \rangle}{\mathbf{e}} \geq \frac{\hat{r}(k)}{\mathbf{e}}.
\end{align*}
This gives, $\hat{r}(k+1) \leq \textstyle \hat{r}(k) - \frac{\hat{\alpha}}{4\mathbf{e}^2}\hat{r}^2(k) + \frac{4\alpha n L_h^2 + L_f +2/\hat{\alpha}}{2} \varepsilon(k)^2.$
% \begin{align}\label{eq:dummy100}
%     \hat{r}(k+1) &\leq \textstyle \hat{r}(k) - \frac{\hat{\alpha}}{2\mathbf{e}^2}\hat{r}^2(k) + \frac{\alpha n L_h^2}{2} \varepsilon^2.
%     % & = \hat{r}(k) - \textstyle \frac{\hat{\alpha}}{2\mathbf{e}^2} \hat{r}^2(k) + \frac{\alpha n L_h^2}{2} \varepsilon^2(k).
% \end{align}
Thus, while $\hat{r}(k)> 2\mathbf{e} \sqrt{\frac{4\alpha n L_h^2 + L_f +2/\hat{\alpha}}{\hat{\alpha}}} \varepsilon_0$, we have,
% $\hat{r}(k+1) \leq \textstyle \hat{r}(k) - \frac{\hat{\alpha}}{4\mathbf{e}^2} \hat{r}^2(k) < \hat{r}(k),$
% Dividing both sides of~(\ref{eq:dummy100}) by $\hat{r}(k+1)\hat{r}(k)$, noting $\hat{r}(k+1) \leq \hat{r}(k)$ and summing up the inequalities we get,
% \begin{align*}
%   \textstyle \frac{1}{\hat{r}(k)} \geq \frac{1}{\hat{r}(0)} + \textstyle \frac{\hat{\alpha} k}{2 \mathbf{e}^2} - \frac{32 \alpha n L_h^2 \mathbf{B}^2}{\delta^2 \hat{r}^2(0) } \sum_{s=0}^{k-1}\lambda^{2k_c(s)} .
% \end{align*}
% Therefore,
% \begin{align*}
%   \hat{r}(k) \leq \frac{2\mathbf{e}^2}{\hat{\alpha}k - \frac{32 \alpha n L_h^2 \mathbf{B}^2}{\delta^2 \hat{r}^2(0)} \sum_{s=0}^{k-1}\lambda^{2k_c(s)} } \leq \frac{2\mathbf{e}^2}{\hat{\alpha}k - \frac{ \alpha n L_h^2 }{2 \hat{r}^2(0)} \overline{\varepsilon}^2} . \hspace{-0.22in} \qed
% \end{align*}
% Further, if $\varepsilon(k)$ is such that $\varepsilon(k) \leq \varepsilon(0) \mu_0^k, k \geq 0$, for some $\mu_0 \in (0,1)$ then,
% \begin{align*}
%   \hat{r}(k) \leq \textstyle \frac{2\mathbf{e}^2}{\hat{\alpha}k - \frac{\alpha n L_h^2}{2 \hat{r}^2(0) c_0}}.
% \end{align*}
% or equivalently, $\hat{r}(k)> \sqrt{2} n L_h \mathbf{e} \varepsilon(k)$, we have 
\begin{align*}
    \hat{r}(k+1) &\leq \textstyle \hat{r}(k) - \frac{\hat{\alpha}}{8\mathbf{e}^2} \hat{r}^2(k) \\
    & \leq \textstyle \hat{r}(k) - \frac{\hat{\alpha}}{8\mathbf{e}^2} \hat{r}(k) 2 n \mathbf{e} \varepsilon_0 L_h \\
    & = \textstyle \left(1-\frac{\alpha}{ 4\mathbf{e}} L_h \varepsilon_0 \right)\hat{r}(k) \\
    & = \beta \hat{r}(k) \\
    \implies \hat{r}(k) &\leq \beta^k \hat{r}(0).  
\end{align*}
Note, $ \alpha L_h \varepsilon_0 < 4 \mathbf{e} \implies \beta \in (0,1)$. This completes the proof. \qed}
\begin{remark}
\textcolor{black}{
Theorem~\ref{thm:conproof} establishes a geometric rate of convergence to a small $O(\varepsilon_0)$ neighborhood of the optimal solution. Since, the gradients of functions $f_i$ are bounded,
\begin{align*}
  f(x^i(k)) - f^* & \leq \hat{r}(k) + \langle \nabla f(x^i(k)),  x^i(k) - \hat{x}(k) \rangle \nonumber\\
  & \hspace{-0.2in} \textstyle \leq \hat{r}(k) + \sum_{j=1}^n \|\nabla f(x^i(k))\|\|x^i(k) - \hat{x}(k)\| \nonumber\\
  & \hspace{-0.2in} \leq \hat{r}(k) + h_m n\varepsilon(k). 
\end{align*}
Hence, using the result of Theorem~\ref{thm:conproof} we conclude that $f(x^i(k)) - f^*$, similar to $\hat{r}(k)$ decreases geometrically, with respect to the outer gradient descent iterations of the Algorithm~\ref{alg:gradcons}, until reaching a small $O(\varepsilon_0)$ neighborhood. The parameter $\varepsilon_0$ is a user specified algorithm parameter which can be chosen appropriately to get solutions arbitrarily close to the optimal solution. Note that the Algorithm~\ref{alg:gradcons} (in the worst case) utilizes $O(k\log k)$ number of communication steps along with $k$ gradient descent iterations. This is a $\log(k)$ factor increase compared to other gradient descent based algorithms in the literature that typically requires $O(k)$ communication steps along with $k$ gradient iterations. The additional communication steps provide improved convergence guarantees for the \texttt{GradConsensus} algorithm over the existing algorithms in the literature. In particular, the Q-linear rate of convergence with respect to the number of gradient computation steps (outer gradient descent iterations of Algorithm~\ref{alg:gradcons}) in Theorem~\ref{thm:conproof} is stronger than the sub-linear convergence rate present in the literature \cite{nedic2009distributed}, \cite{duchi2011dual}, \cite{shi2015extra} under the assumptions~\ref{ass:graph_ass}, \ref{ass:finite_fstar} and~\ref{ass:fi_conv_lip}. Moreover, empirically it is seen that the total number of gradient descent steps is smaller with our algorithm. Thus it is is more suited to situations where computations are expensive.}
\end{remark}
\begin{remark}
We emphasize that even if Assumption~\ref{ass:bounded gradient} is not satisfied the result of Theorem~\ref{thm:conproof} still holds. Using, Lemma~\ref{lem:cons_comm1}, after performing $O(k^2 + k\log(k))$ total number of communication steps (in the worst case) at the $k^{th}$ outer gradient descent iteration the \texttt{GradConsensus} algorithm converges at a Q-linear rate until a neighborhood of the optimal solution.   
\end{remark}
}
\end{subsection}

\begin{subsection}{Convergence Analysis for Strongly Convex $f$}
In this subsection, we make an additional assumption:
\begin{assumption}\label{ass:f_strconv}
$f = \sum_{i=1}^n f_i$ is a strongly convex function with parameter $\sigma > 0$.
\end{assumption}
\noindent Note, that Assumption~\ref{ass:f_strconv} does not require all the $f_i$'s to be necessarily strongly convex. However, when this is the case, assumption~\ref{ass:f_strconv} holds naturally. \textcolor{black}{ \textcolor{black}{We will show in Theorem~\ref{thm:strconvproof} that the solution estimates of every agent converges exactly to the optimal solution of problem~(\ref{eq:introprob}) at a R-linear rate. Similar to Theorem~\ref{thm:conproof} we will provide the total number of communication steps required to achieve the R-linear rate of convergence.}
\begin{theorem}\label{thm:strconvproof}
\textcolor{black}{Let assumptions~\ref{ass:graph_ass}-\ref{ass:fi_conv_lip} and~\ref{ass:f_strconv} hold. Let $\hat{\alpha} \leq \frac{2}{\sigma + L_f}$. Define, $\rho:= \sqrt{1 - \textstyle  \frac{2\hat{\alpha}\sigma L_f}{\sigma + L_f}}$. Let $\varepsilon(k) = \mu^k$, where, $\mu \in [\rho,1)$. Consider, the outer gradient descent iteration $k$, the total consensus communication iterations $\mathcal{K}:= \sum_{s=1}^k k_c(s)$ are bounded by $\sum_{s=0}^k \left[ \frac{\log\big(\frac{1}{\mu^s}\big)}{-\log \lambda} + \frac{\log\big(\frac{8n}{\delta}(\gamma^{s} \frac{\gamma}{\gamma-\mu} + \frac{\gamma^{s}}{nL_h}\alpha L_0) \big)}{-\log \lambda} \right] = O(k^2)$, where, $ L_0:= \max_{1\leq i\leq n} \nabla \|f_i(0)\|,\gamma = 1+\alpha L_h$ and $L_h:= \max_{1 \leq i \leq n} L_{f_i}$. Then at the $k^{th}$ outer gradient descent iteration of Algorithm~\ref{alg:gradcons}, the agent estimates converges at a R-linear rate to the optimal solution,
\begin{align*}
   \|x^i(k) - x^*\| \leq C\mu^k, \ \mbox{for all} \ i \in \mathbf{V},
\end{align*}
where, $C:= \|\hat{x}(0) - x^*\| + \textstyle (2\alpha L_h + 1) \frac{\mu}{\mu-\rho} + 1$.}
\end{theorem}
\noindent \textbf{Proof.}
Consider,
    \begin{align}
        \|\hat{x}(k) - \hat{\alpha}\hat{g}(k) - x^*\|^2 & = \   \|\hat{x}(k) - x^*\|^2  + \hat{\alpha}^2 \|\hat{g}(k)\|^2 \nonumber \\
        & \hspace{0.5in} - 2\hat{\alpha} \langle \hat{g}(k), \hat{x}(k) - x^* \rangle\nonumber \\
        & \hspace{-1.35in} \leq \textstyle \left(1 - \frac{2 \hat{\alpha}}{\sigma + L_f} \right)\|\hat{x}(k) - x^*\|^2 + \textstyle  \hat{\alpha}\left(\hat{\alpha} - \frac{2}{\sigma + L_f} \right)\|\hat{g}(k)\|^2 \nonumber\\
        & \hspace{-1.35in} \leq \rho^2 \|\hat{x}(k) - x^*\|^2\label{eq:dummy_for_ehat}, 
    \end{align}
where, we used Lemma~\ref{lem:strconv_inner_prod_bound} and the fact $\hat{g}(x^*) = 0$. From~(\ref{eq:closedform_upd}),  $ \|\hat{x}(k+1) - x^*\| = \|\hat{x}(k) - \hat{\alpha}g(k) - x^* + v(k)\|  \\
    \leq \|\hat{x}(k) - \hat{\alpha}\hat{g}(k) - x^*\| + \hat{\alpha} \| \hat{g}(k) - g(k) \| + \varepsilon(k).$
Combining~(\ref{eq:dummy_for_ehat}) and the result of Lemma~\ref{lem:gradbound} we get,
\begin{align*}
\|\hat{x}(k+1) - x^*\| &\leq \rho \|\hat{x}(k) - x^*\| + (2\alpha L_h + 1)\varepsilon(k) \\
& = \rho \|\hat{x}(k) - x^*\| + (2\alpha L_h + 1) \mu^k.
\end{align*}
Using the triangle inequality and applying the above inequality recursively we get,
\begin{align*}
    \|x^i(k) - x^*\|  &\leq \|\hat{x}(k) - x^*\|  + \|x^i(k) - \hat{x}(k)\|\\
    & \hspace{-0.45in} \textstyle \leq \rho^k \|\hat{x}(0) - x^*\| + (2\alpha L_h + 1) \sum_{s=0}^{k-1}  \rho^{s} \mu^{k-s} + \mu^k \\
    & \hspace{-0.45in}  \leq \rho^k \|\hat{x}(0) - x^*\| + \textstyle (2\alpha L_h + 1) \mu^k \sum_{s=0}^{k-1} \left(\frac{\rho}{\mu}\right)^s + \mu^k\\
    & \hspace{-0.45in}  \leq \rho^k \|\hat{x}(0) - x^*\| + \textstyle (2\alpha L_h + 1) \mu^k \frac{\mu}{\mu-\rho} + \mu^k\\
    & \hspace{-0.45in}  \leq \mu^k \left[\|\hat{x}(0) - x^*\| + \textstyle (2\alpha L_h + 1) \frac{\mu}{\mu-\rho} + 1\right]\\
    & \hspace{-0.45in} = C\mu^k. \qed
\end{align*}
\begin{remark}
\textcolor{black}{Note that the Algorithm~\ref{alg:gradcons} (in the worst case) utilizes $O(k^2)$ number of communication steps along with $k$ gradient descent iterations for a global R-linear rate of convergence to the exact optimal solution. Using the result in Theorem~\ref{thm:strconvproof} the per node work complexity (sum of total number of gradient computations and total number of communication steps) to achieve an accuracy of $\tau > 0$, i.e., $\|x^i(k) - x^*\| \leq \tau$ is given by: $O(\log(1/\tau)) + O(\log(1/\tau)^2)$. Thus, to obtain an $\eta$-optimal solution, Algorithm~\ref{alg:gradcons} utilizes $O(\log(1/\tau))$ computation steps and $O(\log(1/\tau)^2)$ communication steps.}  
\end{remark}
% \begin{remark}
% Note that the analysis in Theorem~\ref{thm:strconvproof} holds true in the case when Assumption~\ref{ass:bounded gradient} is not satisfied. In particular the proposed \texttt{GradConsensus} algorithm has a global R-linear rate of convergence to the optimal solution. Further, the requirement of the total number of communication steps to be performed is same as in Theorem~\ref{thm:strconvproof} in the worst case. In particular, using, Lemma~\ref{lem:cons_comm1} the total number of communication steps performed at the $k^{th}$ outer gradient descent iteration has a worst case complexity of $O(k^2)$ for achieving a global R-linear rate of convergence.
% \end{remark}
}
\end{subsection}
\textcolor{black}{
\textcolor{black}{Theorems~\ref{thm:conproof} and~\ref{thm:strconvproof} show that utilizing the $\varepsilon$-\texttt{Consensus} protocol also improves the convergence neighborhood of the optimal solution. In particular, the consensus parameter $\varepsilon(k)$ can be suitably chosen to get near-optimal solutions. Theorem~\ref{thm:strconvproof} shows that by controlling the disagreement between the agents (and hence the infeasibility of the algorithm iterates) to a small value using an appropriately chosen $\varepsilon(k)$, results in strong convergence guarantees (global R-linear rate of convergence) for the \texttt{GradConsensus} algorithm. However, there exists a trade-off as tight regulation of the disagreement between agents' estimates would require performing more total number of communication steps $O(k^2)$ until any outer gradient descent iteration $k$, compared to $O(k \log k)$ that gives a Q-linear rate to the near optimal solution.}
}
\end{section}

\begin{section}{Numerical Simulations and Results}\label{sec:simulations}
% \textbf{Graphs and Weight matrix $\mathbf{P}$:}
We consider a network of $100$ agents where the network interconnection topology is generated using the Erdos-Renyi model \cite{erdHos1960evolution} with connectivity probability $0.2$. The weight matrix $\mathbf{P}$ is chosen using the equal neighbor model \cite{olshevsky2009convergence}. We focus on solving the following distributed logistic regression problem,
\begin{align*}
     \minimize_{x \in \mathbb{R}^p} \ \textstyle \sum_{i=1}^n \frac{1}{n_i} \sum_{j = 1}^{n_i} \ln( 1 + \text{exp} (-(A_{ij} x)y_{ij}) ), 
\end{align*}
where each agent $i \in \mathbf{V}$ has its training data $(A_{ij}, y_{ij}) \in \mathbb{R}^p \times \{-1,+1\}$, $j = 1, 2, \dots, n_i$, with feature variables $A_{ij}$ and binary outcomes $y_{ij}$. For our simulations we generate an artificial data-set of feature vectors $A_{ij}$ with outcome $y_{ij} = 1$ from a normal distribution with mean $\mu_1$ and standard deviation $\sigma_1$, and with output $y_{ij} = -1$ from another normal distribution with mean $\mu_2$ and standard deviation $\sigma_2$.
% \begin{align*}
%     \minimize\limits_{x \in \mathbb{R}^{p}} \ \textstyle \frac{1}{2} \sum_{i=1}^{n}\|z_i - H_i x\|^2.
% \end{align*}
% Here, each agent has a measured data $z_i \in \mathbb{R}^{n}$ a measurement scaling matrix  $H_i \in \mathbb{R}^{n \times p}$ of agent $i \in \mathbf{V}$ and $x^{*}$ denotes the optimal solution (unknown signal) that is to be estimated. The entries of the matrices $H_i, i \in \mathbf{V}$, the observed data $z_i$ are chosen from i.i.d samples of standard normal distribution $\mathcal{N}(0,1)$. The unknown signal (or the true state) $x$ is also generated from the i.i.d samples of the standard normal distribution. % For the distributed least squares problem we set $p =10$. \\
% \textbf{Case II - Distributed Logistic Regression:} 
% The distributed logistic regression has the following form:
% \begin{align*}
%      \minimize_{x \in \mathbb{R}^p} \ \sum_{i=1}^n \frac{1}{n_i} \sum_{j = 1}^{n_i} \ln( 1 + \text{exp} (-(A_{ij} x)z_{ij}) ), 
% \end{align*}
% where each agent $i \in \mathbf{V}$ has its training data $(A_{ij}, z_{ij}) \in \mathbb{R}^p \times \{-1,+1\}$, $j = 1, 2, \dots, n_i$, with feature variables $A_{ij}$ and binary outcomes $z_{ij}$. For our simulations we generate an artificial data-set of feature vectors $A_{ij}$ with outcome $z_{ij} = 1$ from a normal distribution with mean $\mu_1$ and standard deviation $\sigma_1$, and with output $z_{ij} = -1$ from another normal distribution with mean $\mu_2$ and standard deviation $\sigma_2$.
% The number of features $p$ are taken to be $15$.
% \textbf{Algorithms to compare:}
We will compare the performance of the proposed \texttt{GradConsensus} algorithm with three state-of-the-art distributed approaches: Distributed Gradient Descent (DGD) \cite{nedic2009distributed},  EXTRA \cite{shi2015extra} and PushPull gradient \cite{pu2020push}. For all the methods we use a constant step-size. \textcolor{black}{In the simulation results demonstrated here, we have chosen a constant $\varepsilon(k) = 0.01, \forall k$ for the \texttt{GradConsensus} algorithm.}
% Each method is initialized with $x^i(0)$ for all $i \in \mathbf{V}$ as a normal random variable.

% \textbf{Simulation Platform:} All the numerical examples in this section are implemented in MATLAB, and run on a desktop computer with 16 GB RAM and an Intel Core i7 processor running at 1.90 GHz.

% \begin{figure}[b]
%     \centering
%   \subfloat{%
%       \includegraphics[width=0.42\linewidth,trim={1.5cm 7cm 2cm 8.2cm},clip]{images/comm_DGD_2.pdf}}
%   \subfloat{%
%         \includegraphics[width=0.42\linewidth,trim={1.5cm 7cm 2cm 8.5cm},clip]{images/comm_EXTRA_2.pdf}}
%     \\
%   \subfloat{%
%         \includegraphics[width=0.42\linewidth,trim={1.5cm 7cm 2cm 8.33cm},clip]{images/comm_PushPull_2.pdf}}
%   \subfloat{%
%         \includegraphics[width=0.42\linewidth,trim={1.5cm 7cm 3cm 8.5cm},clip]{images/comm_GradCons_2.pdf}}
%   \caption{Number of communication steps required by each method in case \textbf{I} for a residual value less than $10^{-5}$.}
%   \label{fig:comm_2} 
% \end{figure}
\begin{figure}[b]
    \centering
     \includegraphics[scale=0.3,trim={1.7cm 6.2cm 1.2cm 6cm},clip] {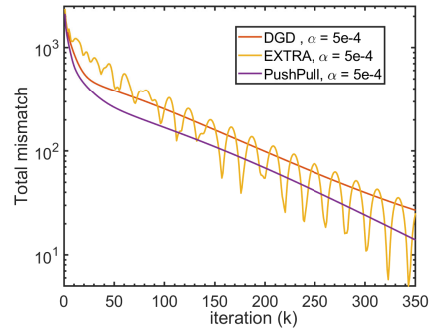}
    \caption{Consensus constraint violation}
    \label{fig:const_violation}
    \vspace{-0.1in}
\end{figure}
\textcolor{black}{\textbf{Consensus Constraint violation:} In Fig.~\ref{fig:const_violation} we plot the total mismatch between the estimates generated by different agents while solving the distributed logistic regression problem for the three algorithms. We calculate the total mismatch at any iteration $k$ as $\frac{\sum_{i=1}^n \sum_{j=1}^n \|x^i(k) - x^j(k)\|}{\sum_{i=1}^n \sum_{j=1}^n \|x^i(0) - x^j(0)\|}$. For all the three algorithms it can be seen that the constraint violation has a significant value. Although, it decreases with the number of iterations but a large number of iterations are required to obtain a small total mismatch among the agents. \textcolor{black}{Hence, any finite-time solution generated by the three algorithms will not provide a viable solution for applications where meeting the constraints is critical. We remark that the proposed \texttt{GradConsensus} algorithm ensures that the agent estimates remain $\varepsilon$-close to each other during each iteration and any finite-time terminated solution results in a viable solution.} 
}

\textbf{Convergence of solution residuals with respect to outer gradient descent iterations:} In the following discussion we present the results of comparison between \texttt{GradConsensus} and other methods in terms of the  solution residual $\frac{\| x^i(k) - x^* \|}{\|x^i(0) - x^*\|}$ with respect to the outer gradient-descent iterations. Here, $i$ is chosen to be the agent that gives the lowest value of the solution residual in the network running the corresponding algorithm.
\begin{figure}[h] 
    \centering
  \subfloat{%
  \hspace{-0.25in}
       \includegraphics[scale=0.24,trim={1.2cm 6.5cm 2cm 6.3cm},clip] {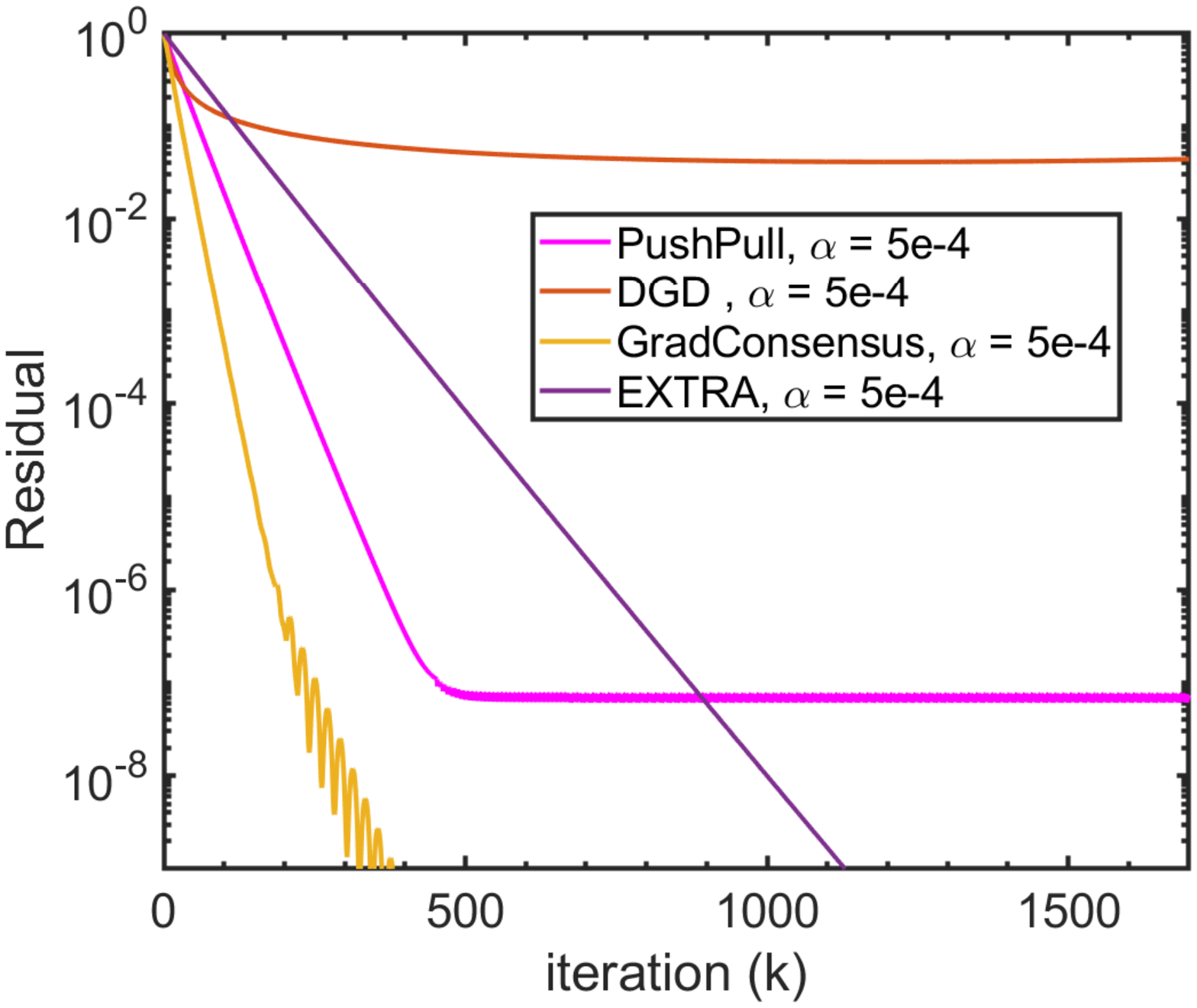}}
  \subfloat{%
    \includegraphics[scale=0.24,trim={1.2cm 6.5cm 1cm 6.2cm},clip] {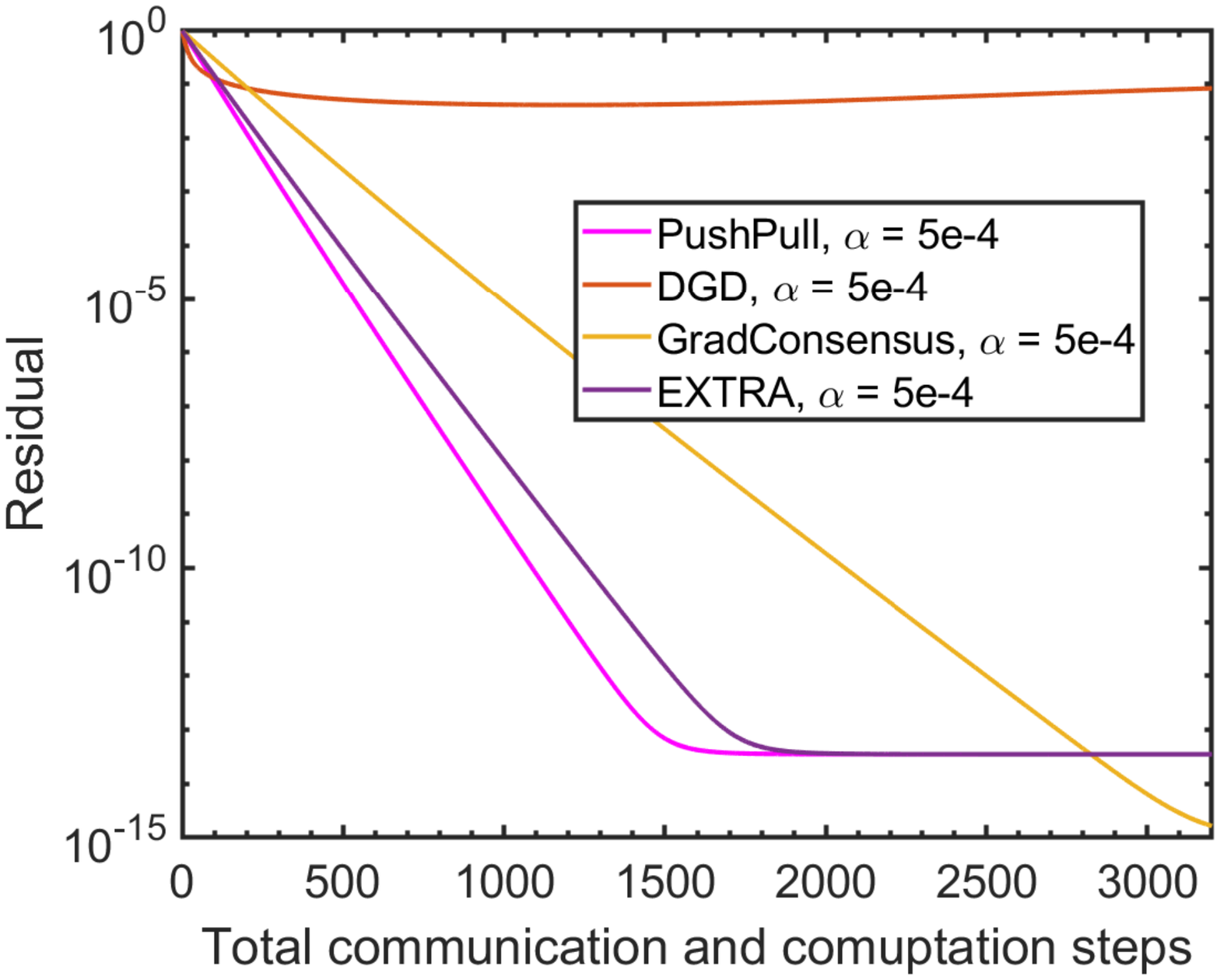}}
  \caption{Residuals $\frac{\| x^i(k) - x^* \|}{\|x^i(0) - x^*\|}$ against the number of outer gradient descent iterations (left) and the total communication and computation steps (right).}
  \label{fig:comparison2_sol} 
\end{figure}
Fig.~\ref{fig:comparison2_sol} gives the residual trajectories for all the compared methods. Note that the \texttt{GradConsensus} gives a superior performance converging to an error less than $10^{-8}$ in approximately $300$ (outer) iterations. Observe that the decrease in the residual is fastest for the \texttt{GradConsensus} algorithm among the compared algorithms, showing applicability of the proposed scheme. EXTRA is the second best method in this case requiring around $1000$ iterations to achieve a similar level of residual error. \textcolor{black}{These results demonstrate that the total number of gradient computation steps required by the proposed \texttt{GradConsensus} algorithm (to reach the same value of solution residual) is significantly less compared to the other algorithms. This shows an advantage of using the \texttt{GradConsensus} algorithm in applications where computation of the gradients is (computation) cost intensive.}

% \begin{figure}[h]
%     \centering
%     \includegraphics[scale=0.35,trim={0cm 6.2cm 1cm 7cm},clip] {images/new_comp_sol_2.pdf}
%     \caption{Residuals $\frac{\| x^i(k) - x^* \|}{\|x^i(0) - x^*\|}$ against the number of iterations for case \textbf{II}.}
%     \label{fig:comparison3_sol}
%     \vspace{-0.1in}
% \end{figure}

% \textbf{Case II:} For case \textbf{II}, we present the trajectories of the residual error for all the four methods in Fig.~\ref{fig:comparison3_sol}. The plots demonstrate that in solving the distributed logistic regression problem \texttt{GradConsensus} algorithm shows superior convergence properties with  converging to an residual error less than $10^{-10}$ in less than $500$ iterations. EXTRA is the second best method with $1200$ iterations for a residual of $10^{-10}$.
% Fig.~\ref{fig:comparison3_fval} gives the convergence paths for the error $f(k) - f^*$ with respect to number of iterations $k$ for case \textbf{II}.
\textbf{Convergence of solution residuals with respect to total number of communication and computation steps:} As, the \texttt{GradConsensus} method utilizes an approximate consensus protocol at each iteration the communication burden of \texttt{GradConsensus} is larger than the other compared methods. To provide a more comprehensive comparison, we compare the convergence of the solution residuals $\frac{\| x^i(k) - x^* \|}{\|x^i(0) - x^*\|}$ for each method with respect to the total number of communication and computation steps required (including the $\varepsilon$-\texttt{Consensus} protocol iterations). Like the previous case the agent $i$ was chosen as the one which gives lowest residual value.
% \begin{figure}[h]
%     \centering
%     \includegraphics[scale=0.35,trim={0cm 6.2cm 1cm 7cm},clip] {images/new_comp_comm_1.pdf}
%     \caption{Residuals $\frac{\| x^i(k) - x^* \|}{\|x^i(0) - x^*\|}$ against the number of communication steps required.}
%     \label{fig:comm1_sol}
%     \vspace{-0.1in}
% \end{figure}
Fig.~\ref{fig:comparison2_sol} gives the residual plots in case of all the four algorithms. We observe that in terms of the communication complexity PushPull is the best performing method in this case requiring around $1000$ communication steps to reach a residual value of less than $10^{-10}$. As discussed (remark 6) due to an $\varepsilon$-\texttt{Consensus} step at each iteration the \texttt{GradConsensus} algorithm takes more number of communication steps to reach the same level of residual (around $2100$ iterations). However, it can be seen in Fig.~\ref{fig:comparison2_sol} that the proposed \texttt{GradConsensus} doesn't stall and converges to a smaller neighborhood of the optimal solution unlike the other algorithms.
% We believe this might be due to a non-strongly convex function instance of the distributed least squares problem. 

% \begin{figure}[h]
%     \centering
%     \includegraphics[scale=0.35,trim={0cm 6.2cm 1cm 7cm},clip] {images/new_comp_comm_2.pdf}
%     \caption{Residuals $\frac{\| x^i(k) - x^* \|}{\|x^i(0) - x^*\|}$ against the number of communication steps required for case \textbf{II}.}
%     \label{fig:comm2_sol}
%     \vspace{-0.1in}
% \end{figure}

% \textbf{Case II:} The plot of residuals with respect to the number of communication steps required in case \textbf{II} is given in Fig.~\ref{fig:comm2_sol}. In this case, PushPull is a better performing method with a residual level of $10^{-13}$ in $1500$ communication steps. EXTRA is the second best algorithm with around $1800$ communication steps for the same level of residual. Note, this result also indicates that the number of communication steps for the \texttt{GradConsensus} algorithm is more compared to other algorithms. However, one advantage of the \texttt{GradConsensus} algorithm is that it converges to a neighborhood of the optimal solution with a lesser value of residual. 
\begin{table}[h]
\scriptsize
    \centering
    \begin{tabular}{|c|c|c|c|}
    \hline
     \multicolumn{4}{|c|}{Table 1: \textbf{Average CPU time of the four methods (sec)}} \\
     \hline
    \texttt{GradConsensus}& EXTRA & DGD& PushPull \\
     \hline 
      2.35 & 3.82 & 5.29 & 3.18\\
     \hline
    \end{tabular}
    \label{tab:cputime}
    % \captionsetup{justification=centering}
    % \caption{Average CPU time of the four methods (sec)}
\end{table}
\textbf{CPU Computation time requirement:} The amount of time required by a processor to execute the instructions of the compared algorithms denotes the CPU time of each algorithm. Table~1 provides average CPU time of the four methods to reach the residual value of less than $10^{-5}$ over $1000$ tests with random graph realizations. Table~1 illustrates that even-though \texttt{GradConsensus} is more communication intensive it takes lesser CPU time to reach a desired solution. The reason for this observation is that the \texttt{GradConsensus} performs significantly less number of computationally expensive gradient computation steps compared to the other methods (as detailed in Fig.~\ref{fig:comparison2_sol}). 
In applications where the computation complexity of the problems is a concern, the \texttt{GradConsensus} method provides a more efficient alternative to the existing methods with faster convergence.

\end{section}
\begin{section}{Conclusion}\label{sec:conclusion}
\textcolor{black}{In this article, we considered the problem of distributively minimizing the sum of $n$ convex functions over a directed multi-agent network. We introduced the \textit{optimize then agree} framework where the optimization step and the consensus step are decoupled to improve the convergence properties of the distributed optimization algorithms by reducing the mismatch between the solution estimates of the agents. We developed a novel \texttt{GradConsensus} algorithm where each agent performs a gradient-descent update for the optimization step and utilizes a finite-time $\varepsilon$-\texttt{Consensus} protocol to achieve $\varepsilon(k)$-close agreement between the agent estimates at each iteration $k$ of the algorithm. 
Further, we established strong convergence guarantees for the proposed \texttt{GradConsensus} algorithm under two different set of assumptions on the aggregate objective function $f$. In particular, we showed that the iterates generated by the \texttt{GradConsensus} algorithm converges to the optimal solution at a linear rate of convergence under these assumptions. In numerical simulations, we applied the \texttt{GradConsensus} to solve the distributed logistic regression problem. The results indicate the suitability of the proposed \texttt{GradConsensus} algorithm in solving distributed optimization problems. We should note that the presented \textit{optimize then agree} framework is applicable to other existing algorithms in the literature and provides a guideline for development of newer algorithms to solve the distributed optimization problem. }
\end{section}

\raggedbottom
\textcolor{black}{
\begin{section}{Appendix}\label{sec:appndx}
\begin{lemma}\label{lem:cons_comm1}
Let Assumptions~\ref{ass:graph_ass} and~\ref{ass:fi_conv_lip} hold. Then at any outer gradient descent iteration $k$ of Algorithm~\ref{alg:gradcons}, after $k_c(k) = \ceil*{
\Big[\frac{\log\big(\frac{1}{\varepsilon(k)}\big)}{-\log \lambda} + \frac{\log\big(\frac{8n}{\delta}(\gamma^{k} \sum_{s=1}^{k-1} \frac{\varepsilon(s)}{\gamma^s} + \frac{\gamma^{k} - 1}{\gamma - 1}\alpha L_0) \big)}{-\log \lambda}\Big]}$, where, $ L_0:= \max_{1\leq i\leq n} \nabla \|f_i(0)\|,\gamma = 1+\alpha L_h$ and $L_h:= \max_{1 \leq i \leq n} L_{f_i}$, iterations of the consensus protocol (updates~(\ref{eq:numerator})-(\ref{eq:ratio})) with the initial condition $s^i(0) = z^i(k), \ t_i(0) = 1, \forall i \in \mathbf{V}$  we have: 
\begin{align*}
 \|x^i(k) - \hat{z}(k)\|  \leq \textstyle \varepsilon(k), \ \forall i \in \mathbf{V},
\end{align*}
where, $\varepsilon(k)$ is the consensus tolerance parameter, $L_h := \max_{1 \leq i \leq n} L_{f_i}$ and $\delta >0$, $\lambda \in (0,1)$ are parameters of the graph $\mathbf{G}$ satisfying $
    \delta \geq \frac{1}{n^n}, \ \ \lambda \leq \textstyle \left(1 - \frac{1}{n^n} \right)$.
\end{lemma}
\textbf{Proof.}
The proof is similar to Lemma~\ref{lem:cons_comm}. In particular, we will utilize~(\ref{eq:nedic_result}) to get an upper bound on the number of consensus iterations $k_c(k)$. Recall, from~(\ref{eq:nedic_result}) that the estimates $x^i(k)$ generated by \texttt{GradConsensus} algorithm at every iteration $k$ converges to the average $\hat{z}(k)$ at a geometric rate. Therefore, we conclude,
\begin{align}\label{eq:nedic_result1}
    \hspace{-0.098in} \|x^i(k) - \hat{z}(k)\| \leq \textstyle \frac{8 \sqrt{n} \lambda^{k_c(k)}}{\delta} \|\z(k)\|, \ \forall i \in \mathbf{V},
\end{align} 
where, $\z (k) := [z^1(k), \dots, z^n(k)] \in \mathbb{R}^{p\times n}$ and, $\delta >0, \lambda \in (0,1)$ satisfy: $
    \delta \geq \frac{1}{n^n}, \ \ \lambda \leq \textstyle \left(1 - \frac{1}{n^n} \right)$.
Here, the variables $\lambda$ and $\delta$ are parameters of the graph $\mathbf{G}$ as defined in Lemma~\ref{lem:cons_comm}. Next, we will bound $\|\z(k)\|$ using an induction argument.\\
\textit{Claim:} At any $k$, 
$\|z^i(k)\| \leq \gamma^{k} \sum_{s=1}^{k-1} \frac{\varepsilon(s)}{\gamma^s} + \frac{\gamma^{k} - 1}{\gamma - 1}\alpha L_0$, for all $i \in \mathbf{V}$, where, $ L_0:= \max_{1\leq i\leq n} \nabla \|f_i(0)\|, \gamma = 1+\alpha L_h$.\\ \textit{Proof:} For $k = 1$, for any $i$ we have,  $\| z^i(1) \| = \| x^i(0) - \alpha \nabla f_i(x^i(0))\| \leq \alpha \max_{1 \leq i \leq n} \nabla \|f_i(x^i(0))\| = \alpha \max_{1 \leq i \leq n} \nabla \|f_i(0)\| = \alpha L_0$. Note, $x^i(0) = 0, \forall i \in \mathbf{V}$ due to the initialization of Algorithm~\ref{alg:gradcons}. Assume, for $k = k, \| z^i(k) \| \leq \gamma^{k} \sum_{s=1}^{k-1} \frac{\varepsilon(s)}{\gamma^s} + \frac{\gamma^{k} - 1}{\gamma - 1}\alpha L_0$, for all $i \in \mathbf{V}$. Now, for $k = k+1$, for any $i$, $\| z^i(k+1) \| = \| x^i(k) - \alpha \nabla f_i(x^i(k))\| \leq \| x^i(k) - \hat{z}(k)\| + \| \hat{z}(k)\| + \alpha \| \nabla f_i(x^i(k))\| \leq \varepsilon(k) + \|\hat{z}(k)\| + \alpha \|\nabla f_i(x^i(k)) - \nabla f_i(x^i(0)) \| + \alpha \|\nabla f_i(x^i(0))\| \leq  \varepsilon(k) + \|\hat{z}(k)\| + \alpha L_h \|x^i(k)\| + \alpha  \|\nabla f_i(0)\| \overset{\mathrm{(\ref{eq:conStep})}}{\leq} \varepsilon(k) + \|\hat{z}(k)\| + \alpha L_h \|\hat{z}(k)\| + \alpha L_h \varepsilon(k) + \alpha L_0 \leq (1+\alpha L_h) [\gamma^{k} \sum_{s=1}^{k-1} \frac{\varepsilon(s)}{\gamma^s} + \frac{\gamma^{k} - 1}{\gamma - 1}\alpha L_0] + (1+\alpha L_h) \varepsilon(k) + \alpha L_0 = \gamma^{k+1} \sum_{s=1}^{k} \frac{\varepsilon(s)}{\gamma^s} + \frac{\gamma^{k+1} - 1}{\gamma - 1}\alpha L_0 $, for all $i \in \mathbf{V}$. Therefore, induction holds. \\
Using the above claim it can be shown that $\|\z(k)\| \leq \sqrt{n}(\gamma^{k} \sum_{s=1}^{k-1} \frac{\varepsilon(s)}{\gamma^s} + \frac{\gamma^{k} - 1}{\gamma - 1}\alpha L_0)$. If $\frac{\varepsilon(k) \delta}{\lambda^k_c(k) 8 \sqrt{n}} = \|\z(k)\| \leq \sqrt{n}(\gamma^{k} \sum_{s=1}^{k-1} \frac{\varepsilon(s)}{\gamma^s} + \frac{\gamma^{k} - 1}{\gamma - 1}\alpha L_0)$ it implies that, \\$\frac{\delta \varepsilon(k)}{8n (\gamma^{k} \sum_{s=1}^{k-1} \frac{\varepsilon(s)}{\gamma^s} + \frac{\gamma^{k} - 1}{\gamma - 1}\alpha L_0)} \leq \lambda^{k_c(k)}$. Therefore, we have,
\begin{align*}
    k_c(k) &\leq \textstyle \frac{-1}{\log \lambda} \Big[\log\big(\frac{1}{\varepsilon(k)}\big) \nonumber \\
    & \textstyle + \log\big(\frac{8n}{\delta}(\gamma^{k} \sum_{s=1}^{k-1} \frac{\varepsilon(s)}{\gamma^s} + \frac{\gamma^{k} - 1}{\gamma - 1}\alpha L_0) \big)\Big] := \overline{k_c}(k).
\end{align*}
Using~(\ref{eq:nedic_result1}) we conclude that after $\overline{k_c}(k)$ number of iterations at the $k^{th}$ outer gradient descent iteration, $\|x^i(k) - \hat{z}(k)\| \leq \varepsilon(k)$, for all $i \in \mathbf{V}$.
\begin{remark}
Lemma~\ref{lem:cons_comm1} provides an upper bound on the number of communication steps required at the $k^{th}$ outer gradient descent iteration of Algorithm~\ref{alg:gradcons} to obtain $\varepsilon(k)$-close solution. In particular, if $0< \varepsilon(k) <1$, then at the $k^{th}$ outer gradient descent iteration of Algorithm~\ref{alg:gradcons}, after $O\big(\log\big(\frac{1}{\varepsilon(k)}\big) + k \big)$ communication steps the estimates of all the agents are guaranteed to be $\varepsilon(k)$-close to each other.  
\end{remark}
\end{section}}

\bibliographystyle{ieeetr} 

\bibliography{references}

\begin{thebibliography}{10}

\bibitem{tsitsiklis1984problems}
J.~N. Tsitsiklis, ``Problems in decentralized decision making and
  computation.,'' tech. rep., Massachusetts Inst of Tech Cambridge Lab for
  Information and Decision Systems, 1984.

\bibitem{bertsekas1989parallel}
D.~P. Bertsekas and J.~N. Tsitsiklis, {\em Parallel and distributed
  computation: numerical methods}, vol.~23.
\newblock Prentice hall Englewood Cliffs, NJ, 1989.

\bibitem{nedic2009distributed}
A.~Nedic and A.~Ozdaglar, ``Distributed subgradient methods for multi-agent
  optimization,'' {\em IEEE Transactions on Automatic Control}, vol.~54, no.~1,
  p.~48, 2009.

\bibitem{lobel2010distributed}
I.~Lobel and A.~Ozdaglar, ``Distributed subgradient methods for convex
  optimization over random networks,'' {\em IEEE Transactions on Automatic
  Control}, vol.~56, no.~6, pp.~1291--1306, 2010.

\bibitem{nedic2010asynchronous}
A.~Nedic, ``Asynchronous broadcast-based convex optimization over a network,''
  {\em IEEE Transactions on Automatic Control}, vol.~56, no.~6, pp.~1337--1351,
  2010.

\bibitem{duchi2011dual}
J.~C. Duchi, A.~Agarwal, and M.~J. Wainwright, ``Dual averaging for distributed
  optimization: Convergence analysis and network scaling,'' {\em IEEE
  Transactions on Automatic control}, vol.~57, no.~3, pp.~592--606, 2011.

\bibitem{chen2012diffusion}
J.~Chen and A.~H. Sayed, ``Diffusion adaptation strategies for distributed
  optimization and learning over networks,'' {\em IEEE Trans. on Signal
  Processing}, vol.~60, no.~8, pp.~4289--4305, 2012.

\bibitem{yuan2016convergence}
K.~Yuan, Q.~Ling, and W.~Yin, ``On the convergence of decentralized gradient
  descent,'' {\em SIAM Journal on Optimization}, vol.~26, no.~3,
  pp.~1835--1854, 2016.

\bibitem{nedic2014distributed}
A.~Nedi{\'c} and A.~Olshevsky, ``Distributed optimization over time-varying
  directed graphs,'' {\em IEEE Transactions on Automatic Control}, vol.~60,
  no.~3, pp.~601--615, 2014.

\bibitem{nedic2016stochastic}
A.~Nedi{\'c} and A.~Olshevsky, ``Stochastic gradient-push for strongly convex
  functions on time-varying directed graphs,'' {\em IEEE Transactions on
  Automatic Control}, vol.~61, no.~12, pp.~3936--3947, 2016.

\bibitem{shi2015extra}
W.~Shi, Q.~Ling, G.~Wu, and W.~Yin, ``Extra: An exact first-order algorithm for
  decentralized consensus optimization,'' {\em SIAM Jour. on Opt.}, vol.~25,
  no.~2, pp.~944--966, 2015.

\bibitem{shi2015proximal}
W.~Shi, Q.~Ling, G.~Wu, and W.~Yin, ``A proximal gradient algorithm for
  decentralized composite optimization,'' {\em IEEE Transactions on Signal
  Processing}, vol.~63, no.~22, pp.~6013--6023, 2015.

\bibitem{li2019decentralized}
Z.~Li, W.~Shi, and M.~Yan, ``A decentralized proximal-gradient method with
  network independent step-sizes and separated convergence rates,'' {\em IEEE
  Transactions on Signal Processing}, vol.~67, no.~17, pp.~4494--4506, 2019.

\bibitem{olshevsky2017linear}
A.~Olshevsky, ``Linear time average consensus and distributed optimization on
  fixed graphs,'' {\em SIAM Journal on Control and Optimization}, vol.~55,
  no.~6, pp.~3990--4014, 2017.

\bibitem{tsianos2012push}
K.~I. Tsianos, S.~Lawlor, and M.~G. Rabbat, ``Push-sum distributed dual
  averaging for convex optimization,'' in {\em IEEE conf. on decision and
  control}, pp.~5453--5458, IEEE, 2012.

\bibitem{kempe2003gossip}
D.~Kempe, A.~Dobra, and J.~Gehrke, ``Gossip-based computation of aggregate
  information,'' in {\em 44th Annual IEEE Symposium on Foundations of Computer
  Science, 2003. Proceedings.}, pp.~482--491, IEEE, 2003.

\bibitem{xin2018linear}
R.~Xin and U.~A. Khan, ``A linear algorithm for optimization over directed
  graphs with geometric convergence,'' {\em IEEE Control Systems Letters},
  vol.~2, no.~3, pp.~315--320, 2018.

\bibitem{pu2020push}
S.~Pu, W.~Shi, J.~Xu, and A.~Nedic, ``Push-pull gradient methods for
  distributed optimization in networks,'' {\em IEEE Transactions on Automatic
  Control}, 2020.

\bibitem{wood2013power}
A.~J. Wood, B.~F. Wollenberg, and G.~B. Shebl{\'e}, {\em Power generation,
  operation, and control}.
\newblock John Wiley \& Sons, 2013.

\bibitem{patel2020distributed}
S.~Patel, B.~Lundstrom, G.~Saraswat, and M.~V. Salapaka, ``Distributed power
  apportioning with early dispatch for ancillary services in renewable grids,''
  {\em arXiv preprint arXiv:2007.11715}, 2020.

\bibitem{Rendezvous_example}
Z.~{Feng} and G.~{Hu}, ``A distributed constrained optimization approach for
  spatiotemporal connectivity-preserving rendezvous of multi-robot systems,''
  in {\em 2018 IEEE Conference on Decision and Control (CDC)}, pp.~987--992,
  2018.

\bibitem{fax2002information}
J.~A. Fax and R.~M. Murray, ``Information flow and cooperative control of
  vehicle formations,'' {\em IFAC Proceedings Volumes}, vol.~35, no.~1,
  pp.~115--120, 2002.

\bibitem{jakovetic2014fast}
D.~Jakoveti{\'c}, J.~Xavier, and J.~M. Moura, ``Fast distributed gradient
  methods,'' {\em IEEE Transactions on Automatic Control}, vol.~59, no.~5,
  pp.~1131--1146, 2014.

\bibitem{chen2012fast}
A.~I.-A. Chen, {\em Fast distributed first-order methods}.
\newblock PhD thesis, Massachusetts Institute of Technology, 2012.

\bibitem{johansson2008subgradient}
B.~Johansson, T.~Keviczky, M.~Johansson, and K.~H. Johansson, ``Subgradient
  methods and consensus algorithms for solving convex optimization problems,''
  in {\em IEEE Conf. on Decision and Control}, pp.~4185--4190, IEEE, 2008.

\bibitem{khatana2020acc}
V.~{Khatana}, G.~{Saraswat}, S.~{Patel}, and M.~V. {Salapaka},
  ``Gradient-consensus method for distributed optimization in directed
  multi-agent networks,'' in {\em 2020 American Control Conference (ACC)},
  pp.~4689--4694, 2020.

\bibitem{berahas2018balancing}
A.~S. Berahas, R.~Bollapragada, N.~S. Keskar, and E.~Wei, ``Balancing
  communication and computation in distributed optimization,'' {\em IEEE
  Transactions on Automatic Control}, vol.~64, no.~8, pp.~3141--3155, 2018.

\bibitem{xiao2004fast}
L.~Xiao and S.~Boyd, ``Fast linear iterations for distributed averaging,'' {\em
  Systems \& Control Letters}, vol.~53, no.~1, pp.~65--78, 2004.

\bibitem{prakash2019distributed}
M.~Prakash, S.~Talukdar, S.~Attree, V.~Yadav, and M.~V. Salapaka, ``Distributed
  stopping criterion for consensus in the presence of delays,'' {\em IEEE
  Transactions on Control of Network Systems}, 2019.

\bibitem{saraswat2019distributed}
G.~Saraswat, V.~Khatana, S.~Patel, and M.~V. Salapaka, ``Distributed
  finite-time termination for consensus algorithm in switching topologies,''
  {\em arXiv:1909.00059}, 2019.

\bibitem{olshevsky2009convergence}
A.~Olshevsky and J.~N. Tsitsiklis, ``Convergence speed in distributed consensus
  and averaging,'' {\em SIAM Journal on Control and Optimization}, vol.~48,
  no.~1, pp.~33--55, 2009.

\bibitem{melbournepreprint}
J.~Melbourne, G.~Saraswat, V.~Khatana, S.~Patel, and M.~V. Salapaka, ``On the
  geometry of consensus algorithms with application to distributed termination
  in higher dimension,'' {\em International Federation of Automatic Control
  (IFAC)}, 2020.

\bibitem{yadav2007distributed}
V.~Yadav and M.~V. Salapaka, ``Distributed protocol for determining when
  averaging consensus is reached,'' in {\em 45th Annual Allerton Conf},
  pp.~715--720, 2007.

\bibitem{prakash2018distributed}
M.~Prakash, S.~Talukdar, S.~Attree, S.~Patel, and M.~V. Salapaka, ``Distributed
  stopping criterion for ratio consensus,'' in {\em 2018 56th Annual Allerton
  Conference on Communication, Control, and Computing (Allerton)},
  pp.~131--135, IEEE, 2018.

\bibitem{qu2019accelerated}
G.~Qu and N.~Li, ``Accelerated distributed nesterov gradient descent,'' {\em
  IEEE Transactions on Automatic Control}, 2019.

\bibitem{nesterov2013introductory}
Y.~Nesterov, {\em Introductory lectures on convex opt.: A basic course},
  vol.~87.
\newblock Springer Science \& Business Media, 2013.

\bibitem{erdHos1960evolution}
P.~Erd{\H{o}}s and A.~R{\'e}nyi, ``On the evolution of random graphs,'' {\em
  Publ. Math. Inst. Hung. Acad. Sci}, vol.~5, no.~1, pp.~17--60, 1960.

\end{thebibliography}

\end{document}